\documentclass[aps,notitlepage, floatfix, jcp]{revtex4-1}%
\usepackage{amssymb}
\usepackage{amsfonts}
\usepackage{amsmath}
\usepackage{graphicx}
\usepackage{enumerate}
\usepackage[usenames,dvipsnames]{color}
\usepackage{ulem}%
\providecommand{\U}[1]{\protect\rule{.1in}{.1in}}
\begin{document}
\title{A Two-parameter Extension of Classical Nucleation Theory}
\author{James F. Lutsko}
\affiliation{Center for Nonlinear Phenomena and Complex Systems, Code Postal 231,
Universit\'{e} Libre de Bruxelles, Blvd. du Triomphe, 1050 Brussels, Belgium}
\email{jlutsko@ulb.ac.be}
\homepage{http://www.lutsko.com}
\author{Miguel A. Dur\'{a}n-Olivencia}
\affiliation{\label{IACT}Laboratorio de Estudios Cristalogr\'{a}ficos. Instituto Andaluz de
Ciencias de la Tierra. CSIC-UGR, Avda. de las Palmeras, 18100 Granada, Spain}
\email{maduran@lec.csic.es}

\begin{abstract}
A two-variable stochastic model for diffusion-limited nucleation is developed
using a formalism derived from fluctuating hydrodynamics. The model is a
direct generalization of the standard Classical Nucleation Theory. The
nucleation rate and pathway are calculated in the weak-noise approximation and
are shown to be in good agreement with direct numerical simulations for the
weak-solution/strong-solution transition in globular proteins. We find that
Classical Nucleation Theory underestimates the time needed for the formation
of a critical cluster by two orders of magnitude and that this discrepancy is
due to the more complex dynamics of the two variable model and not, as often
is assumed, a result of errors in the estimation of the free energy barrier.

\end{abstract}
\date{\today }
\maketitle

\section{Introduction}

The process of nucleation plays an important role in chemistry, materials
science and physics. For conditions of low super-saturation, characterized by
large critical clusters, high energy barriers and, typically, slow (on
laboratory time scales) nucleation rates, classical nucleation theory (CNT)
provides an intuitively appealing framework for answering most questions of
practical interest\cite{Kashchiev}. However, modern applications in
self-assembly, confined geometries and extreme conditions have shifted focus
to the non-classical regime of high supersaturation, small (even nano-scale)
critical clusters, low energy barriers and fast nucleation rates\cite{Tanaka}.
At the same time, and in some cases as a direct result of this shift of
interest, the existence of so-called non-classical nucleation pathways have
become an intense subject of investigation\cite{tWF, VekilovCGDReview2004, LN,
Sear}. Such pathways typically involve multi-step nucleation wherein
meta-stable intermediate phases play a role. Finally, the high-supersaturation
regime is also the case most easily investigated by computer simulation. All
of these motivations combine to give ample practical and theoretical reason to
try to extend our understanding of nucleation to the non-classical regime.

Classical nucleation theory has two important limitations\cite{Kashchiev}. The
first is the use of the capillary approximation for the free energy of a
cluster: the properties of the nucleating phase within the cluster (e.g.,
density, crystallinity, ...) are assumed to be identical to those in the
homogeneous bulk, the interface between this interior of new-phase and the
bath of mother-phase is assumed to have zero width and the only dynamic
property of the cluster is its size (radius, mass, etc.). For the classical
regime, these are generally good approximations since interfacial widths are
typically a few molecular diameters and if a cluster has radius several orders
of magnitude larger than the molecular size, the effect of the finite width of
the radius will be correspondingly small. However, for small clusters,
consisting of perhaps only hundreds of molecules and having radius of only a
few molecular diameters, nearly all molecules feel the effect of the interface
and the assumptions of the capillary approximation become crude at best. Much
more realistic models are available based on finite temperature Density
Functional Theory (DFT)\cite{Evans1979,lutsko:acp} but it is then no longer
possible to characterize a cluster solely by its size.

This leads to a second important limitation of CNT: since the properties of
the material inside the cluster are fixed, the dynamics of cluster formation
can only involve change in the size of a cluster. This is the basis for the
dynamics of CNT wherein one assumes that a cluster can only grow or shrink by
units of a single molecule (thus excluding, e.g., coalescence of clusters),
known as the Becker-D\"{o}ring model\cite{BeckerDoring}. A big part of
nucleation theory is then devoted to the determination of the rates of
molecular attachment and detachment, see e.g. Ref. \onlinecite{Kashchiev}.
Once these are specified, cluster dynamics can be viewed as a stochastic
process and the rate of nucleation then determined using, e.g., Kramer's
theory of barrier crossing\cite{Kramer,Gardiner, Hanggi}.

Recently, a new mesoscopic description of nucleation has been developed based
on fluctuating
hydrodynamics\cite{Lutsko_JCP_2011_Com,Lutsko_JCP_2012_1,Lutsko_JCP_2012_2,Lutsko_JCP_2012_3}%
. In this approach, which we will term Mesoscopic Nucleation Theory (MeNT),
the fundamental quantities are the hydrodynamic fields of density, velocity
and temperature. For the important case of diffusion-limited nucleation, e.g.
of macromolecules in a solvent, this can be reduced to a description solely in
terms of the density. The theory still requires some means for calculating the
free energy of the system, but there is no conceptual limitations preventing
the use of anything from the capillary approximation to very sophisticated DFT
models. Assuming spherical symmetry, as is often but not always done, the
density field describing a cluster can be parameterized and its evolution
described solely in terms of the variation of the
parameters\cite{Lutsko_JCP_2012_1}. When the only parameter is the size of the
cluster, the theory reproduces CNT (specifically, the Zeldovich
equation)\cite{lutsko2013b}. So, this path provides an independent means of
reaching the standard theory that does not require concepts such as monomer
attachment/detachment rates. However, there is nothing that \emph{requires}
restriction to a single parameter: it is also possible to introduce multiple
``order parameters'' such as the density within a cluster, the width of the
interface, etc. in addition to the size thus opening the door to more detailed
and realistic models of cluster formation.

It should be noted that this is certainly not the first attempt to use
multiple order parameters to describe nucleation. First, since as stated
above, nucleation is a problem of barrier crossing and can be related to
Kramer's classic work\cite{Hanggi}, the generalization of Kramer's results to
multiple dimensions (in our language, multiple order parameters) by
Langer\cite{Langer1,Langer2} and others\cite{Gardiner,Hanggi} is of course
relevant. What has previously been lacking was the detailed stochastic model
to which to apply this formalism. There have been several lines of work aimed
at improving the capillary description of the critical cluster, particularly
the predicted excess free energy (i.e. the nucleation barrier). Notably,
Reguerra, Reiss and co-workers developed the Extended Modified Liquid Drop
Model\cite{Reguera2004c} where clusters are characterized by both a mass (or
number of particles) and a volume. Schweitzer, Schmelzer and co-workers
introduced the Generalized Gibbs approach in which critical cluster properties
such as the internal density and temperature are allowed to vary from the
bulk\cite{Schmelzer1987,Schmelzer2006a,Schmelzer2011} with similar aims. Note
that neither of these theories directly addresses the coupling of the
additional parameter to the kinetics. Other developments are based on
DFT-inspired calculations of the free energy of clusters. These are sometimes
described in terms of order parameters and then there has been interest in
using these free energy surfaces to determine nucleation pathways, typically
either via application of additional constraints\cite{Ghosh, Corti,
EvansArcherNucleation, LutskoBubble1} or by gradient descent from the saddle
point (e.g. the critical cluster)\cite{LutskoBubble1,
Lutsko_JCP_2008_3,Philippe,Copolymers,WettingCurved}. The constraint-based
method is simply ad hoc and can lead to highly artificial behavior, see e.g.
Ref.\onlinecite{LutskoBubble1} and, in another context, Zannetti et
al\cite{Zannetti} . The problem with the steepest-descent approaches is that
gradient descent and related methods (nudged elastic band, string method, ...)
always requires a method of calculating distance between two points in the
parameter space\cite{Wales, Lutsko_JCP_2008_3}. Simply having a free energy
surface does not answer this requirement and all applications involve some
sort of ad hoc prescription.

Mesoscopic Nucleation Theory addresses all of these issues in a
self-consistent manner. Given the freedom in choosing the level of description
of the free energy, the properties of the cluster can vary as freely as
desired. Whatever form is chosen, it goes into a dynamical equation from which
nucleation kinetics is derived. There are no a priori assumptions made about
the kinetics: for example, the nucleation pathway need not even pass through
the critical cluster. One obtains a complete description of the nucleation
process from the formation of the initial density fluctuations through to the
deterministic growth of post-critical clusters. Gradient descent plays a role
in the so-called weak noise limit but one uses a measure of distance
\emph{completely specified by the dynamics inherited from the fundamental,
fluctuating hydrodynamics}. This will be made explicit below.

The purpose of this paper is to begin the exploration of such a generalized
approach, which could justifiably be viewed as multi-parameter CNT. It turns
out that there are a number of conceptual problems to be solved, even given a
well-defined formalism, so that attention here is limited to the very simplest
case: condensation of a dense phase from a low-density phase, analogous to
liquid-vapor nucleation. To be specific, we will illustrate our calculations
by applying the formalism to the nucleation of a dense droplet of protein from
a weak solution, a process of intrinsic interest due to its role as the first
step in the non-classical crystallization of
proteins\cite{tWF-Proteins,VekilovCGDReview2004,LN, Gunton2007}. Similarly, in
the interest of simplicity, only one additional order parameter is considered:
the density within the cluster is allowed to vary as well as its size. As will
be shown, this is sufficient to lead to significant conceptual differences in
the nucleation process as described by the classical one-parameter and more
general two-parameter theories. We will quickly be led to construct the model
so as to impose mass conservation leading to a cluster structure very similar
to that of the Modified Extended Liquid Drop model mentioned above. Our
primary practical conclusion is that the differences between CNT and our more
complex model cannot be explained merely by the differences in the calculated
free energy barriers, as is often assumed, but rather are also due to the
difference in dynamics of the two theories.

In Section II we outline the theoretical framework that will be used to
construct our extended description of nucleation and we show how CNT can be
easily derived within this context. The next Section is devoted to the
development of a two-variable model. The naive extension of CNT is shown to
fail and the additional concept of mass conservation is invoked to construct a
physically acceptable model. In Section IV, we apply the theory to the
description of the weak-solution/dense-solution transition in globular
proteins. In particular, we discuss the systematic differences observed
between CNT and the two-variable theory. We summarize our conclusions in
Section V.

\section{Theory}

Our goal is the description of diffusion limited nucleation as is applicable
to phase transitions in colloids and solutions of macromolecules. The
prototypical problem is the nucleation of a dense phase - either a disordered,
liquid-like droplet or an ordered crystalline phase - from a weak solution.
However, the theory is equally applicable to the reverse - i.e. the nucleation
of a low-density phase from a dense solution (the analog of the formation of a
gas bubble in a one-component liquid).

The fundamental viewpoint of MeNT is that the new phase forms a cluster which
is characterized by its space- and time-dependent local density (e.g. the
density of the colloid or macromolecule which is equivalent to its
concentration in the solution). A disordered phase has a density which is
(nearly) constant while an ordered phase has a density field exhibiting
molecular-scale structure. In either case, the evolution of the colloidal
density is the focus of attention. Since the initial, low-concentration
solution is presumed to be a metastable phase, there is an energy barrier that
must be overcome for nucleation to occur and the theory must describe the
thermal fluctuations which are responsible for driving the system over this barrier.

The density field is in general coupled to other fields, e.g. temperature and
velocity. However, we specialize here to the case of diffusion limited
nucleation as is appropriate for large molecules (or colloids) in solution.
The system is, in consequence of the coupling to the bath, athermal.
Furthermore, we assume over-damping which allows the velocity field to be
eliminated as described in Ref. \onlinecite{Lutsko_JCP_2012_1}.

\subsection{Framework}

In the over-damped regime, the density field, $\rho\left(  \mathbf{r}\right)
$, obeys a Langevin equation of the
form\cite{Lutsko_JCP_2011_Com,Lutsko_JCP_2012_1}%
\begin{equation}
\frac{d}{dt}\rho\left(  \mathbf{r};t\right)  =D\mathbf{\nabla}\cdot\rho\left(
\mathbf{r};t\right)  \mathbf{\nabla}\left(  \frac{\delta\beta F\left[
\rho\right]  }{\delta\rho\left(  \mathbf{r}\right)  }\right)  _{\rho\left(
\mathbf{r}\right)  \rightarrow\rho\left(  \mathbf{r};t\right)  }%
+\mathbf{\nabla}\cdot\varepsilon\sqrt{2D\rho\left(  \mathbf{r};t\right)  }%
\xi\left(  \mathbf{r};t\right)
\end{equation}
where $D$ is the low-density diffusion constant, $F\left[  \rho\right]  $ is a
coarse-grained free energy functional, $\beta=1/k_{B}T$ is the inverse
temperature and $\varepsilon$ is an artificial parameter included for
explanatory purposes. The physical case always corresponds to $\varepsilon=1$.
The term $\xi\left(  \mathbf{r};t\right)  $ is a fluctuating force that is
white and delta-correlated in both space and time. The diffusion constant is
related to the temperature via $D=k_{B}T/\gamma m$ where $\gamma$
characterizes the friction experienced by the large molecules due to the bath
and where $m$ is the mass of the large molecules. While it is possible to work
directly with the density field, here we simplify the problem by introducing a
restricted set of variables. To do this, the density is represented in terms
of a collection of $N$ parameters as
\begin{equation}
\rho\left(  \mathbf{r};t\right)  =f\left(  \mathbf{r};x^{1}\left(  t\right)
,x^{2}\left(  t\right)  ,...x^{N}\left(  t\right)  \right)  \equiv\rho\left(
\mathbf{r};\mathbf{x}\left(  t\right)  \right)
\end{equation}
for some fixed functional form $f$. (The second equivalence defines a
short-hand notation for this representation.) For example, to describe a dense
cluster, the form of $f$ might be a sigmoidal function with the parameters
being the position and width of the interface. We note that such parameters
are in general constrained for physical reasons:\ for example, the radius and
density both must be non-negative. This $N$-dimensional parameter space will
be denoted $\Sigma$. For an under-saturated system, we expect that the
equilibrium state will be described by a particular set of parameters
corresponding to a spatially constant density and that the dynamics, in the
absence of thermal fluctuations, will tend to drive any deviation from this
state towards this equilibrium state. In a supersaturated system, this will be
true for a certain region of parameter space - the metastable region - but
there will be other parts of parameter space corresponding to super-critical
clusters which will tend (in the absence of thermal fluctuations) to grow
indefinitely thus converting the entire system to the new phase.

It has been shown\cite{Lutsko_JCP_2012_1} that the stochastic differential
equation for the field gives rise to an approximate dynamics for the
parameters. In the particular case of an infinite system with spherical
symmetry, the parameters obey
\begin{equation}
\frac{dx^{i}}{dt}=-Dg^{ij}\left(  \mathbf{x}\right)  \frac{\partial\beta
\Omega}{\partial x^{j}}+D\varepsilon^{2}A^{i}\left(  \mathbf{x}\right)
+\varepsilon\sqrt{2D}q_{a}^{i}\left(  \mathbf{x}\right)  \xi^{a}\left(
t\right)  \label{SDE}%
\end{equation}
where the Einstein summation convention is used and this equation must be
interpreted using the Ito calculus\cite{Lutsko_JCP_2012_1}. The matrix of
kinetic coefficients $g^{ij}\left(  \mathbf{x}\right)  $ is related to the
amplitude of the noise via $g^{ij}\left(  \mathbf{x}\right)  =q_{a}^{i}\left(
\mathbf{x}\right)  q_{a}^{j}\left(  \mathbf{x}\right)  $. In the following, we
use standard covariant notation whereby $g^{ij}$ is the inverse of the matrix
$g_{ij}$ so that $g^{ij}g_{jk}=\delta_{k}^{i}$, the latter being the usual
Kronecker delta function. As discussed elsewhere, the matrix $g_{ij}$ plays
the role of a "metric" in the space of parameters, $\mathbf{x}$ and this will
be relevant when we discuss numerical simulation of these equations (see, e.g.
Appendices \ref{Simulations} and \ref{AveragePaths}). For a given
parameterization of the density, and in three dimensions, it (the inverse of
the matrix of kinetic coefficients) is calculated as%
\begin{equation}
g_{ij}\left(  \mathbf{x}\right)  =\int_{0}^{\infty}\frac{1}{4\pi r^{2}%
\rho\left(  r;\mathbf{x}\right)  }\frac{\partial m\left(  r;\mathbf{x}\right)
}{\partial x^{i}}\frac{\partial m\left(  r;\mathbf{x}\right)  }{\partial
x^{j}}dr \label{g}%
\end{equation}
where the cumulative mass density is%
\begin{equation}
m\left(  r;\mathbf{x}\right)  =4\pi\int_{0}^{r}\rho\left(  r^{\prime
};\mathbf{x}\right)  r^{\prime2}dr^{\prime}.
\end{equation}
The other quantities appearing in Eq.(\ref{SDE}) are the grand-potential,%
\begin{equation}
\Omega\left(  \mathbf{x}\right)  =F\left(  \mathbf{x}\right)  -\mu N\left(
\mathbf{x}\right)
\end{equation}
where $F\left(  \mathbf{x}\right)  $ is just the coarse-grained free energy,
$F\left[  \rho\right]  $, evaluated using the model density $\rho\left(
\mathbf{r};\mathbf{x}\left(  t\right)  \right)  $ and the total number of
particles is $N\left(  \mathbf{x}\right)  =\lim_{r\rightarrow\infty}m\left(
r;\mathbf{x}\right)  $. (Technically, this quantity is infinite but we can
equally well work with the difference from the background which is always
finite.) The noise term is white with correlations $\left\langle \xi
_{i}\left(  t\right)  \xi_{j}\left(  t^{\prime}\right)  \right\rangle
=\delta_{ij}\delta\left(  t-t^{\prime}\right)  $ and the noise amplitude is
related to the metric by $q^{i}_{a}q^{j}_{b}\delta^{ab}=g^{ij}$ (e.g., see
Appendix \ref{MatrixQ} for the explicit relation with two parameters). The
final new term appearing in the SDE is the anomalous force. Before giving its
form, we note that the SDE for the order parameters is not Ito-Stratonovich
equivalent. When it is written in the Ito form, the anomalous force takes the
form
\begin{equation}
A^{i}=\frac{\partial g^{ij}}{\partial x_{j}}+\frac{1}{2}g^{ij}g^{lk}%
\frac{\partial g_{lk}}{\partial x_{j}}+\Delta A^{i} \label{A}%
\end{equation}
The term $\Delta A^{i}$ is given elsewhere\cite{Lutsko_JCP_2012_1} and will be
neglected here for several reasons. First, it vanishes identically for a
single order parameter and its inclusion would complicate the comparison
between the one- and two-variable theories. Second, it also does not occur in
the case of an infinite number of parameters, i.e. in the full hydrodynamic
theory. This leads us to suspect it is an artifact of the approximations
required in the derivation. Finally, the structure of the theory without this
term is much nicer allowing for several important analytic results as
discussed below. The effect of including it will be the subject of a later investigation.

The Fokker-Plank equation describing the probability to observe parameters
$\mathbf{x}$ at time $t$, $P(\mathbf{x},t)$, follows directly from the
stochastic model, Eq.(\ref{SDE}), and is%
\begin{equation}
\frac{\partial}{\partial t}P=-D\frac{\partial}{\partial x^{i}}\left\{
-g^{ij}\left(  \mathbf{x}\right)  \frac{\partial\beta\Omega}{\partial x^{j}%
}+\varepsilon^{2}A^{i}\left(  \mathbf{x}\right)  -\varepsilon^{2}%
\frac{\partial}{\partial x^{j}}g^{ij}\right\}  P
\end{equation}
It is immediately apparent that the stationary probability density (valid only
for under-saturated solutions) is simply
\begin{equation}
P\left(  \mathbf{x}\right)  =\mathcal{N}\sqrt{\det g\left(  \mathbf{x}\right)
}e^{-\beta\Omega\left(  \mathbf{x}\right)  } \label{stationary}%
\end{equation}
where $\mathbf{x}$ is any collection of parameters (i.e. in the simplest case,
it could be the radius or excess mass of the cluster), $\mathcal{N}$ is a
normalization constant and $det g(\mathbf{x})$ will here and hereafter always
denote the determinant of the inverse of the matrix of kinetic coefficients
(i.e. of the metric).

Nucleation occurs when the system starts somewhere in the metastable region
and is driven by thermal fluctuations to the stable region. We characterize
the (inverse of the) nucleation rate by the mean-first passage time for the
formation of a critical cluster \cite{Hanggi}. For the one dimensional case,
there is an exact expression for this quantity (discussed below), however, in
the general case no such result exists. The standard
result\cite{Langer1,Talkner,Hanggi,Gardiner} valid in the weak noise limit is,
in our language,%
\begin{equation}
t_{mfp}=\varepsilon^{-1}\frac{\pi}{D\left\vert \lambda_{-}\right\vert }%
\frac{\sqrt{\left\vert \det\beta\Omega_{ij}^{\left(  c\right)  }\right\vert }%
}{\sqrt{\det g(\mathbf{x}_{c})}\left(  2\pi\right)  ^{N/2}}\left(
\int_{\Sigma_{meta}}\sqrt{\det g\left(  \mathbf{x}\right)  }e^{-\beta
\Omega\left(  \mathbf{x}\right)  }d\mathbf{x}\right)  e^{\beta\Omega
(\mathbf{x}_{c})}\label{T}%
\end{equation}
where $\Omega_{ij}^{\left(  c\right)  }$ is the Hessian of the free energy
evaluated at the critical cluster $\mathbf{x}_{c}$, $\lambda_{-}$ the (sole)
negative eigenvalue of $g^{ij}(\mathbf{x}_{c})\Omega_{jk}^{\left(  c\right)
}$ and $N$ is the number of order parameters. The critical cluster is
determined as usual by $\left.  \partial\Omega(\mathbf{x})/\partial
x^{i}\right\vert _{\mathbf{x}_{c}}=0$.

\subsection{A comment on covariance}

It is common, in a heuristic context, to suppose that the probability of
observing a cluster of mass $M$ is $P\left(  M\right)  \sim e^{-\beta
\Delta\Omega\left(  M\right)  }dM$ where $\Delta\Omega\left(  M\right)  $ is
the work of formation of the cluster or, in other contexts, to suppose that
the probability to observe a cluster of radius $R$ is $\widetilde{P}\left(
R\right)  \sim e^{-\beta\Delta\widetilde{\Omega}\left(  R\right)  }dR$. Since
the two quantities are related by the cluster's density, $M=\frac{4\pi}%
{3}R^{3}\rho$, the work of formation in the two cases is related by
$\Delta\widetilde{\Omega}\left(  R\right)  =\Delta\Omega\left(  M\left(
R\right)  \right)  =\Delta\Omega\left(  \frac{4\pi}{3}R^{3}\rho\right)  $ and
the probability densities are related unambiguously by $P\left(  M\right)
dM=\widetilde{P}\left(  R\right)  dR$ or $\widetilde{P}\left(  R\right)
=\frac{dM}{dR}P\left(  M\left(  R\right)  \right)  =4\pi R^{2}\rho P\left(
M\left(  R\right)  \right)  $. Hence, it cannot be simultaneously be true that
both probability densities are given by $e^{-\beta\Delta\Omega}$. This
ambiguity is termed a lack of covariance, by which we mean that the result
seems to depend on which variable we choose to start with. Nature, of course,
is unambiguous and such a dilemma indicates that some element of physics is missing.

In MeNT, no such ambiguity is present. For example, in the under-saturated
state, the probability density to observe a cluster described by the variables
$\mathbf{x}$ is $P\left(  \mathbf{x}\right)  \mathbf{=}\mathcal{N}\sqrt{\det
g\left(  \mathbf{x}\right)  }e^{-\beta\Delta\Omega\left(  \mathbf{x}\right)
}$. For some other set of variables, $\mathbf{y}$, related to the original
variables as $\mathbf{y}(\mathbf{x})$, we get $\widetilde{P}\left(
\mathbf{y}\right)  \mathbf{=}\widetilde{\mathcal{N}}\sqrt{\det\widetilde{g}%
\left(  \mathbf{y}\right)  }e^{-\beta\Delta\widetilde{\Omega}\left(
\mathbf{y}\right)  }$ where of course $\Delta\widetilde{\Omega}\left(
\mathbf{y}\right)  =\Delta\Omega\left(  \mathbf{x}\left(  \mathbf{y}\right)
\right)  $. However, from the definition given above, it is immediately
apparent that $\sqrt{\det\widetilde{g}\left(  \mathbf{y}\right)  }=\sqrt{\det
g\left(  \mathbf{x}\left(  \mathbf{y}\right)  \right)  }\left\vert
\frac{\partial\mathbf{x}}{\partial\mathbf{y}}\right\vert $ where the second
factor is the Jacobian of the change of variables. Hence, we see that
\begin{align}
\widetilde{P}\left(  \mathbf{y}\right)  d\mathbf{y}  &  \mathbf{=}%
\widetilde{\mathcal{N}}\sqrt{\det\widetilde{g}\left(  \mathbf{y}\right)
}e^{-\beta\Delta\widetilde{\Omega}\left(  \mathbf{y}\right)  }d\mathbf{y}\\
&  =\widetilde{\mathcal{N}}\sqrt{\det g\left(  \mathbf{x}\left(
\mathbf{y}\right)  \right)  }\left\vert \frac{\partial\mathbf{x}}%
{\partial\mathbf{y}}\right\vert e^{-\beta\Delta\Omega\left(  \mathbf{x}\left(
\mathbf{y}\right)  \right)  }d\mathbf{y}\nonumber\\
&  =\widetilde{\mathcal{N}}\sqrt{\det g\left(  \mathbf{x}\left(
\mathbf{y}\right)  \right)  }e^{-\beta\Delta\Omega\left(  \mathbf{x}\left(
\mathbf{y}\right)  \right)  }d\mathbf{x}\nonumber\\
&  =P\left(  \mathbf{x}\right)  d\mathbf{x}\nonumber
\end{align}
so that the result is independent of which set of equivalent variables one
starts with. Similarly, the expression for the mean first passage time given
above can also be seen to be covariant. This serves to confirm the internal
consistency of the theory.

\subsection{Classical Nucleation Theory}

Classical Nucleation Theory (CNT) fits easily within this framework. Suppose
we wish to describe the nucleation of a new phase with density (or
concentration) $\rho_{0}$ in a system with initial density $\rho_{\infty}$.
The notation is motivated by the idea that the density is a spatially varying
field in the underlying fluctuating hydrodynamics and a cluster will have the
density or the new phase, $\rho_{0}$, near the origin and that of the old
phase, $\rho_{\infty}$, far away from the origin. We suppose  that when the
density is constant (i.e. in the bulk state) we are able to calculate the free
energy per unit volume as a function of average density, $\omega\left(
\rho\right)  $, and the planar surface tension at coexistence, $\gamma^{(c)}$.
Note that the former depends on both the temperature and the chemical
potential, $\mu$ and can be expressed in terms of the Helmholtz free energy
per unit volume, $f\left(  \rho\right)  $, as $\omega\left(  \rho\right)
=f\left(  \rho\right)  -\mu\rho$. In terms of these quantities, the excess
free energy (relative to the background or mother phase) of a cluster of
radius $R$ is calculated using the capillary approximation as%
\begin{equation}
\Delta\Omega\left(  R;T,\mu\right)  =V\left(  R\right)  \left(  \omega\left(
\rho_{0}\right)  -\omega\left(  \rho_{\infty}\right)  \right)  +S\left(
R\right)  \gamma^{(c)} \label{FE}%
\end{equation}
where the $V\left(  R\right)  $ and $S\left(  R\right)  $ are the volume and
surface area, respectively. The bulk densities satisfy, for fixed $\mu$ and
$T$, $f^{\prime}\left(  \rho_{0}\right)  =\mu=f^{\prime}\left(  \rho
_{0}\right)  $. Implicit in this model is a density profile whereby the local
density as a function of distance from the center of the cluster, $\rho\left(
r\right)  $, is piece-wise constant having $\rho(r) = \rho_{0}$ for $r<R$ and
$\rho(r) = \rho_{\infty}$for $r>R$. The radius varies with time and, as such,
is the single order parameter characterizing the cluster. A simple calculation
gives%
\begin{equation}
g_{RR}=4\pi\frac{\left(  \rho_{0}-\rho_{\infty}\right)  ^{2}}{\rho_{\infty}%
}R^{3},A=-\frac{3}{8\pi R^{4}}\frac{\rho_{\infty}}{\left(  \rho_{\infty}%
-\rho_{0}\right)  ^{2}}%
\end{equation}
so that the time evolution of the radius is
\begin{equation}
\frac{dR}{dt}=-D\frac{\rho_{\infty}}{4\pi\left(  \rho_{0}-\rho_{\infty
}\right)  ^{2}R^{3}}\frac{\partial\beta\Omega}{\partial R}-D\frac{3}{8\pi
}\frac{\rho_{\infty}}{\left(  \rho_{\infty}-\rho_{0}\right)  ^{2}}%
R^{-4}+\varepsilon\sqrt{2D\frac{\rho_{\infty}}{4\pi\left(  \rho_{0}%
-\rho_{\infty}\right)  ^{2}R^{3}}}\xi\left(  t\right)
\end{equation}
and the Fokker-Planck equation is
\begin{equation}
\label{FP-CNT}\frac{\partial}{\partial t}P=-D\frac{\rho_{\infty}}{4\pi\left(
\rho_{0}-\rho_{\infty}\right)  ^{2}}\frac{\partial}{\partial R}\left\{
-R^{-3}\frac{\partial\beta\Omega}{\partial R}-D\frac{3}{2}R^{-4}%
-\frac{\partial}{\partial R}R^{-3}\right\}  P
\end{equation}
For under saturated conditions, equilibrium is achieved in the uniform state
with density $\rho_{\infty}$ and radius $R=0$. The distribution for
fluctuations in cluster size is%
\begin{equation}
P\left(  R;T,\mu\right)  =\mathcal{N}R^{3/2}e^{-\beta\Delta\Omega\left(
R;T,\mu\right)  } \label{PR_CNT}%
\end{equation}
where $\mathcal{N}$ is determined by normalization. In the limit of large
clusters, the $R^{-4}$ term in Eq.(\ref{FP-CNT}) above can be neglected and
the result is equivalent to the Zeldovich equation of
CNT\cite{Kashchiev,lutsko2013b}. In this way, we see that CNT is recovered
within the formalism of MeNT.

For a supersaturated system, the free energy has a minima at the metastable
density $\rho_{\infty}$ with $R=0$ and at the stable density $\rho_{0}$ with
$R\rightarrow\infty$. There is as well a free energy maximum at the critical
radius,
\begin{equation}
\label{Rcnt}R_{c}=\frac{2\gamma^{(c)}}{\omega\left(  \rho_{\infty}\right)
-\omega\left(  \rho_{0}\right)  }%
\end{equation}
with%
\begin{equation}
\label{Ecnt}\Delta\Omega\left(  R_{c};T,\mu\right)  =\frac{16\pi}{3}%
\frac{\gamma^{(c)3}}{\left(  \omega\left(  \rho_{\infty}\right)
-\omega\left(  \rho_{0}\right)  \right)  ^{2}}.
\end{equation}
In this one-dimensional case, the mean first passage time for a cluster that
is initially in the metastable state to cross the nucleation barrier is
\cite{Gardiner}%
\begin{equation}
\label{mfpt_exact}t_{mfp}= D^{-1} \frac{4\pi\left(  \rho_{0}-\rho_{\infty
}\right)  ^{2}}{\rho_{\infty}}\int_{0}^{R_{c}}dR\;R^{3/2}e^{\beta\Delta
\Omega\left(  R\right)  }\;\int_{0}^{R}dR^{\prime}R^{\prime3/2}e^{-\beta
\Delta\Omega\left(  R^{\prime}\right)  }\;
\end{equation}
An approximate evaluation is possible by first noting that the factor
$e^{\beta\Delta\Omega\left(  R\right)  }$ is large near $R=R_{c}$ and that the
second integral is slowly varying for $R$ near $R_{c}$ since its largest
contribution comes from the neighborhood of $R^{\prime}=0$. Hence, the upper
limit of the second integral can be replaced by $R_{c}$ giving%
\begin{equation}
t_{mfp}\simeq D^{-1} \frac{4\pi\left(  \rho_{0}-\rho_{\infty}\right)  ^{2}%
}{\rho_{\infty}}\int_{0}^{R_{c}}dR\;R^{3/2}e^{\beta\Delta\Omega\left(
R\right)  }\;\int_{0}^{R_{c}}dR^{\prime}R^{\prime3/2}e^{-\beta\Delta
\Omega\left(  R^{\prime}\right)  } \label{Tprod}%
\end{equation}
Then, quadratic approximations of the free energy near the points $R=0$ and
$R=R_{c}$ give the approximate evaluation
\begin{equation}
t_{mfp}\simeq D^{-1} \frac{4\pi\left(  \rho_{0}-\rho_{\infty}\right)  ^{2}%
}{\rho_{\infty}}\left(  R_{c}^{3/2}\sqrt{\frac{\pi}{2\left\vert \beta
\Delta\Omega\left(  R_{c}\right)  \right\vert }}\right)  \;\left(  \left(
\beta\Delta\Omega^{\prime\prime}\left(  0\right)  \right)  ^{-5/2}\frac{\pi
}{2^{\frac{5}{4}}\Gamma\left(  \frac{3}{4}\right)  }\right)  \;
\end{equation}
or%
\begin{equation}
t_{mfp}\simeq D^{-1} \frac{\left(  \rho_{0}-\rho_{\infty}\right)  ^{3/2}}%
{\rho_{0}}\frac{\pi^{3/4}}{4\sqrt{2}\Gamma\left(  \frac{3}{4}\right)
}\left\vert \Delta\omega\right\vert ^{-3/2}\gamma^{(c)-\frac{1}{4}}%
e^{\beta\Delta\Omega\left(  R_{c}\right)  }\;\;
\end{equation}
It is interesting to compare these successive approximations with the general
approximate result given above, Eq.(\ref{T}), which here reduces to%
\begin{align}
t_{mfp}  &  = \sqrt{\frac{\pi g_{RR}\left(  R_{c}\right)  }{2D \Delta
\Omega^{\prime\prime}\left(  R_{c}\right)  }}\left(  \int_{0}^{R_{c}}%
\sqrt{g_{RR}\left(  R\right)  }e^{-\beta\Omega\left(  R\right)  }dR\right)
e^{\beta\Omega\left(  R_{c}\right)  }\\
&  = D^{-1}\frac{4\pi\left(  \rho_{0}-\rho_{\infty}\right)  ^{2}}{\rho
_{\infty}}\sqrt{\frac{\pi}{2\Delta\Omega^{\prime\prime}\left(  R_{c}\right)
}}R_{c}^{3/2}\left(  \int_{0}^{R_{c}}R^{3/2}e^{-\beta\Omega\left(  R\right)
}dR\right)  e^{\beta\Omega\left(  R_{c}\right)  }\nonumber
\end{align}
This corresponds to the saddle point approximation for the first integral
in\ Eq.(\ref{Tprod}) and an exact evaluation of the second.

\section{Two-parameter models}

\subsection{A naive extension of CNT}

\subsubsection{The density profile}

Our goal is to generalize the CNT description of nucleation. It is well-known
that real clusters do not form with an interior density equal to the bulk
value for the new state as assumed in CNT (see, e.g. the simulation results of
ten Wolde and Frenkel\cite{frenkel_gas_liquid_nucleation} as well as DFT
calculations such as Ref. \onlinecite{Lutsko_JCP_2008_3}). We will therefore
attempt to allow the interior density of the cluster to vary as well as the
radius which means treating the cluster density, $\rho_{0}$ in Eq.(\ref{FE})
above, as a variable. Of course, the final state will be one with infinite
radius and the equilibrium bulk density $\rho_{0}^{(bulk)}$ determined by
$\omega^{\prime}(\rho_{0}^{(bulk)})=0$. The kinetic coefficients now form a
symmetric $2 \times2$ matrix. The elements of its inverse (the metric) are
calculated from Eq.(\ref{g}) with the result
\begin{align}
g_{RR}  &  =4\pi\frac{\left(  \rho_{0}-\rho_{\infty}\right)  ^{2}}%
{\rho_{\infty}}R^{3}\\
g_{R\rho_{0}}  &  =\frac{4\pi}{3}\frac{\left(  \rho_{0}-\rho_{\infty}\right)
}{\rho_{\infty}}R^{4}\nonumber\\
g_{\rho_{0}\rho_{0}}  &  =\frac{4\pi}{45} \left(  \frac{1}{\rho_{0}}+\frac
{5}{\rho_{\infty} } \right)  R^{5}\nonumber
\end{align}
and $\det g = \frac{\rho_{\infty}}{5\rho_{0}} \left(  \frac{4 \pi}{3}
\frac{\rho_{0}-\rho_{\infty}}{\rho_{\infty}}R^{4} \right)  ^{2}$.

\subsubsection{Excess free energy}

This generalization immediately raises an issue with the free energy because
one expects, on physical grounds, that two density distributions which are the
same should have the same free energy. However, even if one sets $\rho_{0}
=\rho_{\infty}$, the free energy is not zero unless $R=0$.\ In fact, once the
interior density can change, one must take account of the fact that the
interfacial energy depends on the difference in the densities of the interior
and exterior phases. More microscopic squared-gradient theories suggest that
this energy should be proportional to $\left(  \rho_{0}-\rho_{\infty}\right)
^{2}$ (see Appendix \ref{SurfaceTension}). We therefore use
\begin{equation}
\Delta\Omega\left(  R,\rho_{0}\right)  =V\left(  R\right)  \left(
\omega\left(  \rho_{0}\right)  -\omega\left(  \rho_{\infty}\right)  \right)
+S\left(  R\right)  K\left(  {\rho_{0}-\rho_{\infty}}\right)  ^{2}%
\end{equation}
where $\rho_{\infty}^{\left(  c\right)  }$and $\rho_{0}^{\left(  c\right)  }$
are the equilibrium bulk densities at coexistence and $K\equiv\gamma^{\left(
c\right)  }/\left(  \rho_{0}^{\left(  c\right)  }-\rho_{\infty}^{\left(
c\right)  }\right)  ^{2}$.

\subsubsection{Fluctuations in the under-saturated state}

Once the density profile and free energy are known, the equilibrium
probability density describing fluctuations of the order parameters is given
explicitly by Eq.(\ref{stationary}) which becomes
\begin{equation}
P\left(  \rho_{0},R\right)  =\mathcal{N}\frac{4\pi}{3\sqrt{5}}\frac{\left\vert
\rho_{0}-\rho_{\infty}\right\vert }{\sqrt{\rho_{0}\rho_{\infty}}}%
R^{4}e^{-\beta\Delta\Omega(R,\rho_{0})}%
\end{equation}
where $\mathcal{N}$ is the normalization factor. For sufficiently
under-saturated systems, one expects that the density fluctuations are small
and that we can expand the density-dependence of the distribution in terms of
$\Delta\rho\equiv\rho_{0}-\rho_{\infty}$ giving%
\begin{equation}
P\left(  R,\rho\right)  =\mathcal{N}\frac{4\pi}{3\sqrt{5}}\frac{1}%
{\rho_{\infty}}\left\vert \Delta\rho\right\vert R^{4}\exp\left(  -\beta\left(
V\left(  R\right)  \omega^{\prime\prime}\left(  \rho_{\infty}\right)
+KS\left(  R\right)  \right)  (\Delta\rho)^{2}\right)  \left(  1+O\left(
\Delta\rho\right)  \right)
\end{equation}
The marginal probability density for the radius is%
\begin{equation}
P\left(  R\right)  \sim\int_{0}^{\infty}P\left(  R,\rho_{0}\right)  d\rho
_{0}\sim\mathcal{N}\frac{4\pi}{3\sqrt{5}}\frac{1}{\rho_{\infty}}\frac{R^{4}%
}{\beta V\left(  R\right)  \omega^{\prime\prime}\left(  \rho_{\infty}\right)
+\beta KS\left(  R\right)  }%
\end{equation}
A grave problem is now apparent as this grows as $R$ for large $R$ and so that
the probability density is not normalizable: this simple, intuitive extension
of CNT is not even able to describe the under-saturated, equilibrium state
much less the process of nucleation.

\subsection{Two parameter model with mass conservation}

\subsubsection{Density profile}

The difficulty with the stationary distribution arises because there is
essentially no cost to the formation of a cluster except for that due to the
free energy. In reality however, there is also a dynamic "cost" in that mass
must be transported from one part of the system to another in order to form
any type of density fluctuation. This suggests that a density profile
including mass conservation might give a more realistic description of the system.

There are two types of processes that move mass. The first is diffusion, which
takes place slowly, on long timescales, compared to the second process,
thermal fluctuations. Fluctuations, represented by the noise term in the
stochastic models, must conserve mass but have no intrinsic time scale. Since
they occur too fast for diffusion to be important, any increase in mass in one
region must simultaneously be compensated by a decrease elsewhere. At the
level of fluctuating hydrodynamics, this is strictly enforced by the fact that
the fluctuating force involves a spatial gradient: any increase in mass at
position $r$ is compensated by a corresponding decrease at $r \pm dr$.

In the capillary model, we cannot enforce mass conservation at this local
level, so we will insist that any increase in mass in the core of the cluster
be compensated by a decrease in a region outside the cluster. The structure of
a cluster is therefore generalized to include three regions: an inner core of
radius $R_{0}$ and density $\rho_{0}$, a middle region of radius $R_{1} >
R_{0}$ and density $\rho_{1}$ and the surrounding mother phase with density
$\rho_{\infty}$,
\begin{equation}
\rho\left(  r\right)  =\rho_{0}\Theta\left(  R_{0}-r\right)  +\rho_{1}%
\Theta\left(  r-R_{0}\right)  \Theta\left(  R_{1}-r\right)  +\rho_{\infty
}\Theta\left(  r-R_{1}\right)  .
\end{equation}
The density $\rho_{1}$ is fixed by the requirement that the total mass is
conserved,
\begin{equation}
V_{0}\rho_{0}+\left(  V_{1}-V_{0}\right)  \rho_{1}=V_{1}\rho_{\infty}%
\end{equation}
where $V_{0} \equiv V(R_{0})$, etc. There are now two interfaces: that between
the inner core region and the intermediate region and that between the
intermediate region and the mother phase so the free energy becomes
\begin{equation}
\label{Emassconserving}\Delta\Omega=\Delta\omega(\rho_{0}) V_{0}+ \Delta
\omega(\rho_{1}) \left(  V_{1}-V_{0}\right)  +K\left(  \rho_{1}-\rho
_{0}\right)  ^{2}S_{0}+K\left(  \rho_{\infty}-\rho_{1}\right)  ^{2}S_{1}%
\end{equation}
or, after a little simplification,%
\begin{equation}
\Delta\Omega=\Delta f(\rho_{0}) V_{0}+\Delta f(\rho_{1}) \left(  V_{1}%
-V_{0}\right)  +K\left(  \rho_{0}-\rho_{\infty}\right)  ^{2}\frac{V_{1}%
^{2}S_{0}+V_{0}^{2}S_{1}}{\left(  V_{0}-V_{1}\right)  ^{2}}%
\end{equation}
where we note that the chemical potential drops out (hence the replacement
$\omega\rightarrow f$) due to the mass conservation. For a cluster with higher
core density than the mother phase, the intermediate region will have a lower
density than the background and therefore represents a depletion zone outside
the cluster. In the less intuitive case that the core region has lower density
than the background, the intermediate region will have a higher than average
density, an ``augmentation'' zone. Clearly, the mass of the inner zone is
constrained by the available material in the depletion zone.

We must still specify the radius $R_{1}$. Within this model, the mass to form
a cluster of radius $R_{0}$ is being borrowed from the shell $R_{0} < r <
R_{1}$. As stated above, in the underlying fluctuating hydrodynamics, this
``borrowing'' is strictly local and it was the fact that we were allowing
$R_{1} \rightarrow\infty$ in the first attempt to extend CNT that led to the
divergence noted above. A first guess might therefore to be to impose a fixed
width for the depletion zone by setting $R_{1} = \Delta R_{10} + R_{0}$ for
some constant $\Delta R_{10}$. However, this does not allow clusters to grow
indefinitely: for example, as a super-critical cluster of given density,
$\rho_{0}$, increases in radius, eventually this condition would force
$\rho_{1}(R_{0}) \rightarrow0$ at which point further growth is not possible.
Of course, in reality, steady growth is fed by diffusion which here must be
modeled by an outer radius that grows sufficiently fast as the cluster
accumulates mass. In order to understand the problem of having sufficient mass
for a cluster to grow, we note that the outer density will tend to a constant
as $R$ increases and for a given inner density, $\rho_{0}$ if $R_{1}^{3} =
\Delta R^{3}_{10} + (\rho_{0}/\rho_{\infty})R^{3}_{0}$. Hence, the minimal
model that allows for clusters of arbitrary size has the form $R_{1}^{3} =
\Delta R^{3}_{10} + \lambda R^{3}_{0}$ with constant $\lambda$ chosen ``large enough''.

Unfortunately, this is still not adequate: clusters can now grow to arbitrary
size but it is precisely the fact that \emph{low density} clusters could grow
arbitrarily large that led to the lack of normalization of the distribution.
It is therefore the case that any choice for $R_{1}(R_{0})$ that grows
sufficiently fast as to allow for arbitrary sized clusters gives an
un-normalizable equilibrium distribution (see Appendix \ref{Metric}). The only
solution is to allow the outer radius to depend on density as well, and in
such a way that $R_{1}$ is bounded for small-amplitude fluctuations: for the
case above, we need $\lambda\rightarrow0$ as $\rho_{0} \rightarrow\rho
_{\infty}$. Heuristically, this seems consistent with the idea that small
deviations from the background, i.e. \textit{fluctuations}, can only borrow
mass over a finite region whereas larger deviations from the background, which
are typically the result of multiple fluctuations over a period of time, can
benefit from diffusive spreading of the depletion zone. This then suggests the
model we will investigate which is
\begin{equation}
R_{1}^{3} = \Delta R^{3}_{10} + \lambda\left(  \frac{\rho_{0}-\rho_{\infty}%
}{\rho_{\infty}} \right)  ^{2}R^{3}_{0}%
\end{equation}
where $\Delta R_{10}$ and $\lambda$ are constants and the squared density is
used so that small fluctuations above and below the background are equally
likely. (We could equally well use the first power and take an absolute value,
but that would lead to analytic difficulties that we prefer to avoid.) Using
this, the kinetic coefficients can be evaluated as above (see Appendix
\ref{Metric}).

\subsubsection{Fluctuations in the sub-critical state}

The equilibrium distribution in the sub-critical state is given by
Eq.(\ref{stationary}). Expanding the determinant of the kinetic coefficients
to second order, see Appendix \ref{Metric}, gives the probability density
\begin{equation}
P(R,\rho)=\mathcal{N}\frac{4\pi}{15}\frac{\left\vert \rho_{0}-\rho_{\infty
}\right\vert }{\rho_{\infty}}R^{4}\left(  \frac{5+6y+3y^{2}+y^{3}}{\left(
1-y\right)  \left(  1+y+y^{2}\right)  ^{4}}\right)  ^{1/2}e^{-\beta
\Delta\Omega(R_{0},\rho_{0})}\left(  1+O\left(  \frac{\rho_{0}-\rho_{\infty}%
}{\rho_{\infty}}\right)  ^{2}\right)  \Theta\left(  \Delta R_{10}-R\right)
,\,\,y\equiv R/\Delta R_{10}%
\end{equation}
and using the small-fluctuation expansion of the free energy%
\begin{equation}
\Delta\Omega=\frac{4\pi R^{2}}{\left(  1-y^{3}\right)  ^{2}}\left(  \frac
{1}{6}R\left(  1-y^{3}\right)  \rho_{\infty}^{2}f^{\prime\prime}\left(
\rho_{\infty}\right)  +K\rho_{\infty}^{2}\left(  1+y^{4}\right)  \right)
\left(  \frac{\rho_{0}-\rho_{\infty}}{\rho_{\infty}}\right)  ^{2}\left(
1+O\left(  \frac{\rho_{0}-\rho_{\infty}}{\rho_{\infty}}\right)  \right)
\end{equation}
gives the marginal probability density for the radius
\begin{equation}
P\left(  R\right)  \simeq\mathcal{N}\frac{2}{5\rho_{\infty}}R^{2}\frac{\left(
1-y\right)  ^{3/2}\left(  5+6y+3y^{2}+y^{3}\right)  ^{1/2}}{\left(
1-y^{3}\right)  R\beta f^{\prime\prime}\left(  \rho_{\infty}\right)  +6\beta
K\left(  1+y^{4}\right)  }\Theta\left(  \Delta R_{10}-R\right)  \label{AppR}%
\end{equation}
The fact that the radius is bounded means that normalization is assured. We
remark on the fact that this is clearly a non-Gaussian distribution that looks
nothing like the form assumed in CNT, see Eq.(\ref{PR_CNT}).

\section{Numerical Results}

We now examine a specific application of this model in order to investigate
the implications of the two-variable description of nucleation versus the
usual one-variable CNT.

\subsection{Application to Globular Proteins}

\begin{figure}
[ptb]\includegraphics[angle=0,scale=0.4]{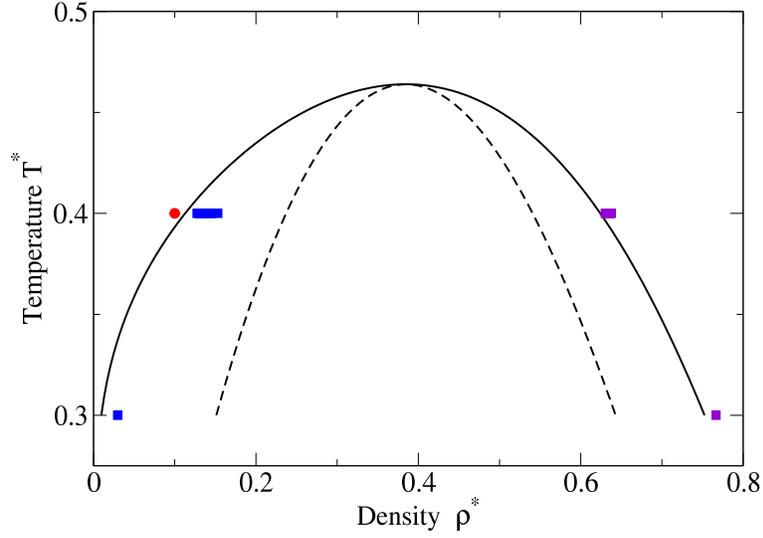}
\caption{Low density/High density solution phase diagram for the tWF potential used in this work. The solid lines are the liquid-liquid binodal and the broken lines are the spinodal. The points indicate stable (circle) and metastable (squares) systems discussed in the text. For the (weak solution) metastable systems, the corresponding stable (dense liquid) states are also indicated.}
\label{fig_eos}
\end{figure}

\begin{figure}
[ptb]\includegraphics[angle=0,scale=0.7]{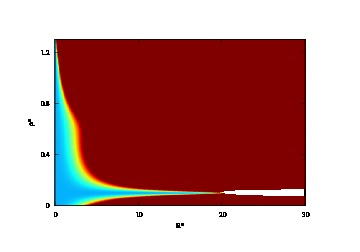}
\includegraphics[angle=0,scale=0.7]{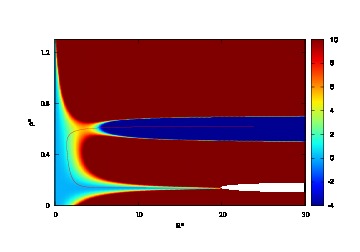}
\caption{Free energy surfaces for the system in a sub-critical (left panel, $\rho_{\infty}^{*} = 0.10$) and super-critical (right panel, $\rho_{\infty}^{*} = 0.1375$) state for the model with parameters $\lambda = 10$ and $\Delta R_{10}^{*} = 20$ and for $T^{*} = 0.4$. For display purposes, he free energy values shown are truncated so that what is shown is $\min(10,\max(-4,\beta \Delta \Omega))$. Both panels show similar extended  ``equilibrium'' basins ($0 \le \beta \Delta \Omega \lesssim 1$, light blue)  and physically forbidden regions (white) which the model cannot access. The super-critical system has a critical point with energy $\beta \Delta \Omega = 5$ and also shows the stable basin at large radii and densities (dark blue): in reality, the free energy drops indefinitely with increasing radius. The right panel also shows the most likely path for nucleation (red line, as calculated in the weak noise limit, see discussion in Section \ref{pathway}). }
\label{fig_fe_surface}
\end{figure}

The two-parameter models constructed above require only the bulk equation of
state and the surface-tension parameter as inputs. Our model system consists
of globular proteins in solution, for which the assumption of
diffusion-limited dynamics is reasonable. In this Section, we present
numerical calculations based on an equation of state derived from the ten
Wolde-Frenkel model pair potential for globular proteins\cite{tWF-Proteins},
\begin{equation}
v\left(  r\right)  =\left\{
\begin{array}
[c]{c}%
\infty,\;\;r\leq\sigma\\
\,\frac{4\,\epsilon}{\alpha^{2}}\left(  \,\left(  \frac{1}{(\frac{r}{\sigma
})^{2}-1}\right)  ^{6}-\,\alpha\,\left(  \frac{1}{(\frac{r}{\sigma})^{2}%
-1}\right)  ^{3}\right)  ,\;\;r\geq\sigma
\end{array}
\right.
\end{equation}
using the standard value $\alpha=50$ and cutoff at $r_{c}=2.5\sigma$ and
shifted so that $v\left(  r_{c}\right)  =0$. The parameter $\sigma$
corresponds to the physical size of the globular protein molecule and
$\epsilon$ could be determined by fitting to the second-virial coefficient in
a weak solution. In the following, numerical results will always be reported
in reduced units using $\sigma$ and $\epsilon$ to scale lengths and energies
respectively. Reduced quantities will be marked with an asterisk, e.g. reduced
temperature is $T^{*} \equiv k_{B}T/\epsilon$ where $k_{B}$ is Boltzmann's
constant. The equation of state was calculated using thermodynamic
perturbation theory with a hard-sphere reference state as discussed in Ref.
[\onlinecite{Lutsko2011a}]. Figure \ref{fig_eos} shows the phase diagram for
the low-concentration/high-concentration liquid phases (equivalent to a
vapor-liquid phase diagram for a one component system). The surface-tension at
coexistence (as required for the free energy model) was calculated using a
previously derived approximation, also based only on the
pair-potential\cite{Lutsko2011a}. Figure \ref{fig_fe_surface} shows the free
energy surfaces for the system in a sub-critical and a super-critical state.
The equilibrium basin is composed of two regions: one with large radius and
small deviations of the density from the background (long-wavelength
fluctuations), the other having small radius and significant deviations from
the background (small, dense, unstable clusters).

\begin{figure}
[ptb]\includegraphics[angle=0,scale=0.4]{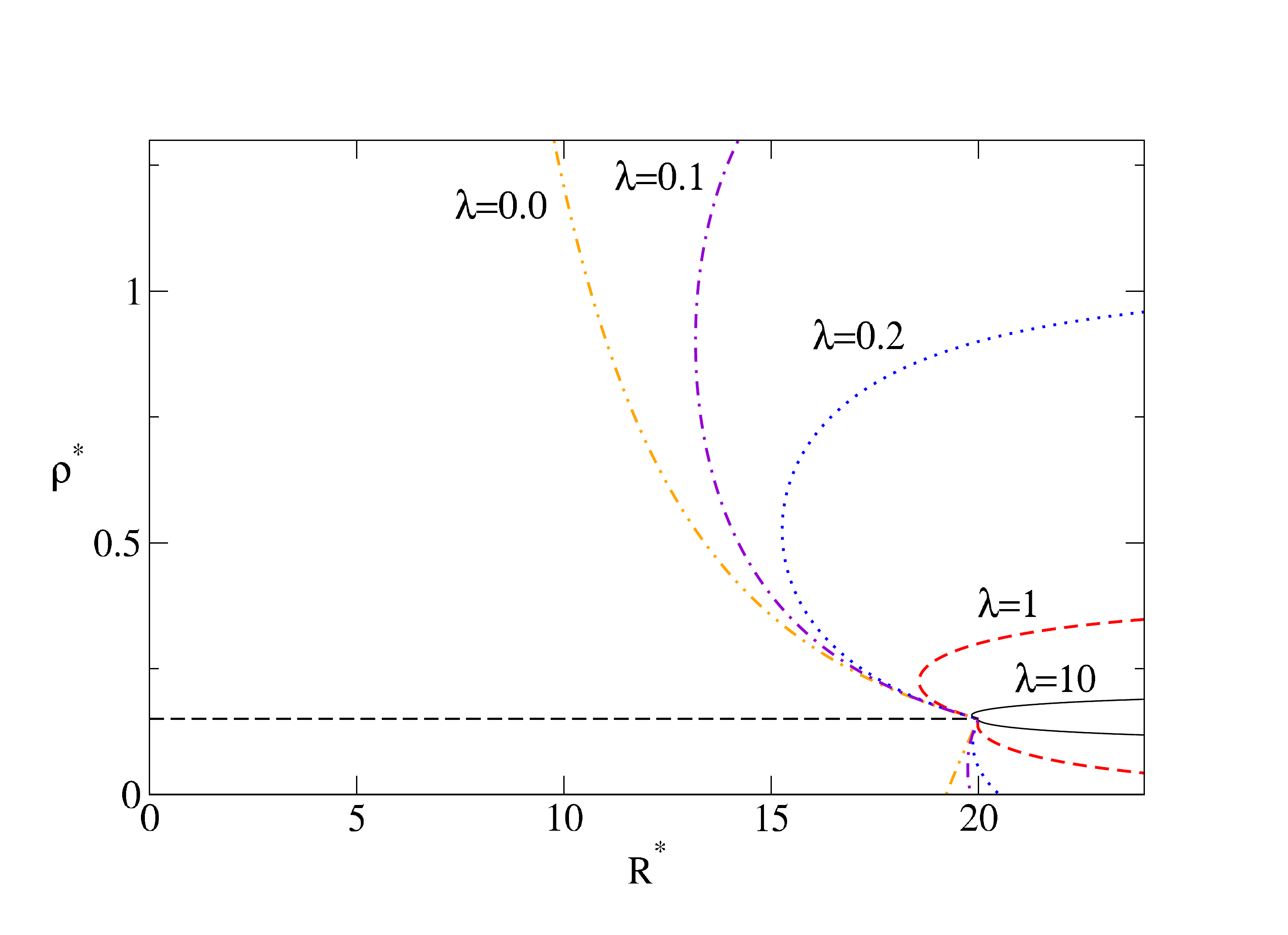}
\caption{The accessible parameter space for a background density $\rho_{\infty}^{*} =0.15$. The parameter $\Delta R_{10}^{*}$ is fixed at $20$ and the effect of different values of $\lambda$ is shown. In each case, the area to the right and between the two branches is inaccessible. For large values of $\lambda$ this reduces to a narrow region around the background density having radius greater than $\Delta R_{10}$.  }
\label{restricted}
\end{figure}

The effect of the parameters $\Delta R_{10}$ and $\lambda$ are illustrated in
Fig. \ref{restricted}. When $\lambda= 0$, there is a fixed, finite amount of
matter available to build a cluster and only a part of the parameter space is
accessible. This is equivalent to a finite system and could be used to study
nucleation under this constraint. As $\lambda$ increases, allowing for growth
of the depletion region with growing cluster size, more of the parameter space
becomes accessible until for $\lambda= 10$ nearly all of the parameter space
is accessible except for a small region near the background region having
radius greater than $\Delta R_{10}$. In all cases, $\Delta R_{10}$ limits the
size of small amplitude clusters.

\subsection{The sub-critical equilibrium state}

\begin{figure}
[ptb]\includegraphics[angle=0,scale=0.5]{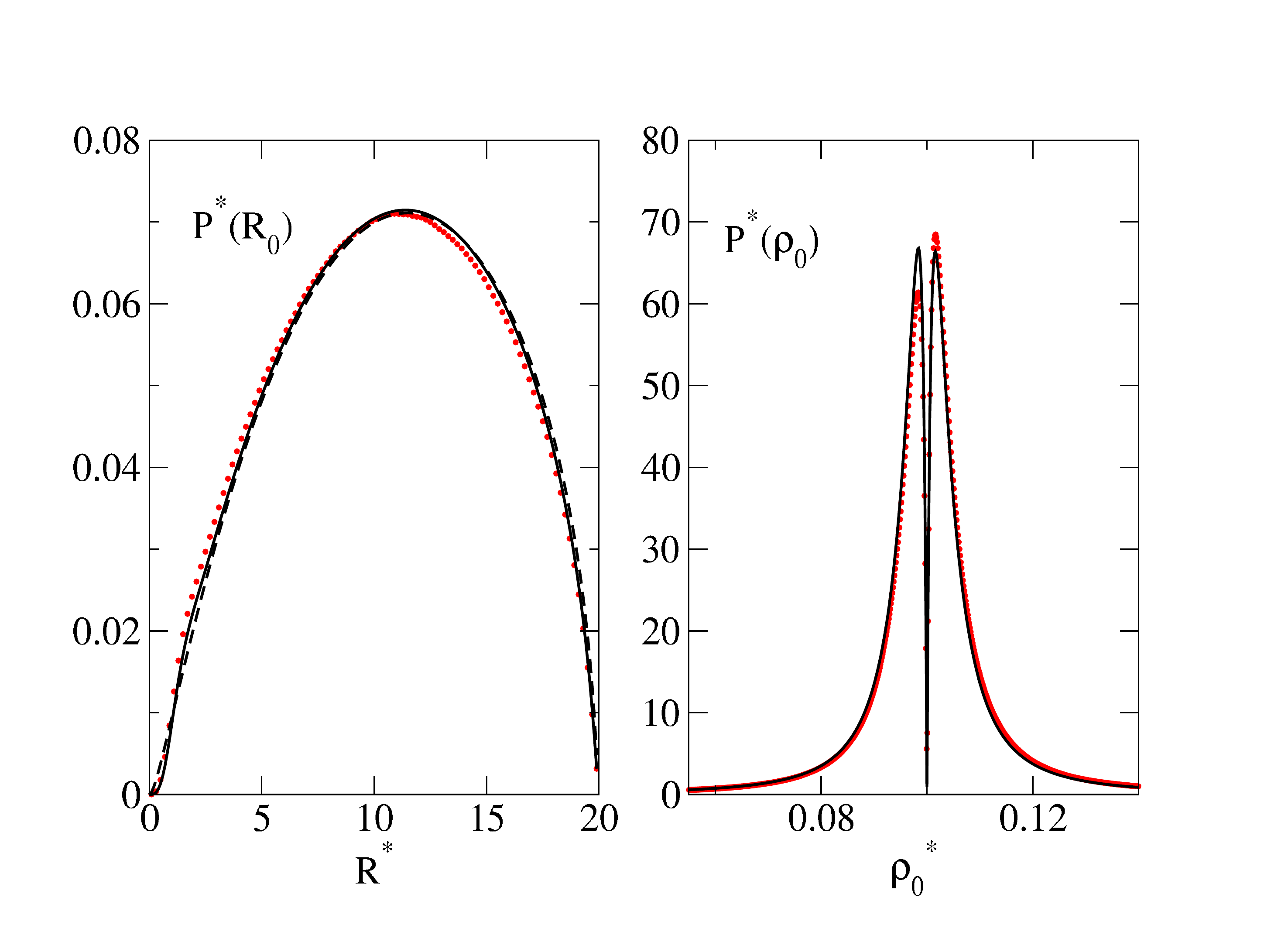}
\caption{Marginal probability densities for the radius (left panel) and the central density (right panel) of clusters for a sub-critical system. The points were determined using histograms constructed from numerical simulation of the stochastic differential equations. The dashed line is the approximation given in Eq.(\ref{AppR}). The state conditions were $T^{*}=0.4$, $\lambda^{*} = 20$ and $\rho^{*}_{\infty}=0.10$.}
\label{fig_comparison}
\end{figure}

One of our main goals in this Section is to compare our analytic
approximations to direct numerical simulation of the stochastic model, Eq.
\ref{SDE}. The stochastic simulations were carried out using a Milstein scheme
with a variable time-step\cite{SDE0,SDE1,SDE2} as described in Appendix
\ref{Simulations}. Figure \ref{fig_comparison} shows the marginal probability
density for the radius and density as determined by numerical integration of
the joint distribution, Eq.(\ref{stationary}) evaluated using the explicit
forms for the free energy, Eq.(\ref{FE}) and for the determinant of the matrix
of kinetic coefficients, Eq.(\ref{gbig})-(\ref{gderiv}), and by simulation of
the stochastic differential equations for a period of $t^{*} = 5 \times10^{6}%
$. Very good numerical agreement is found and further comparisons have shown
that the agreement varies systematically with the quality control of the
stochastic simulations (see the discussion in Appendix \ref{Simulations}). The
analytic approximation for the marginal probability density for the radius,
Eq.(\ref{AppR}) is also shown in the Figure and is clearly a very good approximation.

\subsection{The Nucleation Pathway}

\label{pathway} With more than one order parameter, there are many ways to
transition from the meta-stable to the stable state. In fact, there is not
even an a priori guarantee that the typical nucleation pathway will pass
through (or near) the critical cluster. Nevertheless, it can be
shown\cite{Lutsko_JCP_2011_Com,Lutsko_JCP_2012_1} that given the particular
structure of the stochastic model used here and working in the weak noise
limit, the \textit{most likely path} (MLP) does indeed pass through the
critical cluster and that it can be determined by starting at the critical
cluster, perturbing slightly and integrating the deterministic force,
\begin{equation}
\frac{dx^{i}}{dt}=-Dg^{ij}\frac{\partial\beta\Omega}{\partial x^{j}}.
\label{MLP}%
\end{equation}
The separatrix is the line between the basin of attraction of the two
metastable states. It can be calculated in a similar manner by reversing the
sign of the gradient of the free energy in Eq.(\ref{MLP}), perturbing slightly
in the direction of the stable eigenvector and integrating numerically.

%In essence, the weak noise limit is appropriate provided the thermodynamic
%forces, e.g. the gradient of the free energy, are typically large (more
%precisely, we need that the dimensionless quantities $\sqrt{D}q^{ij}%
%\frac{\partial\beta\Omega}{\partial x^{j}}>1$).

\begin{figure}
[ptb]\includegraphics[angle=0,scale=0.29]{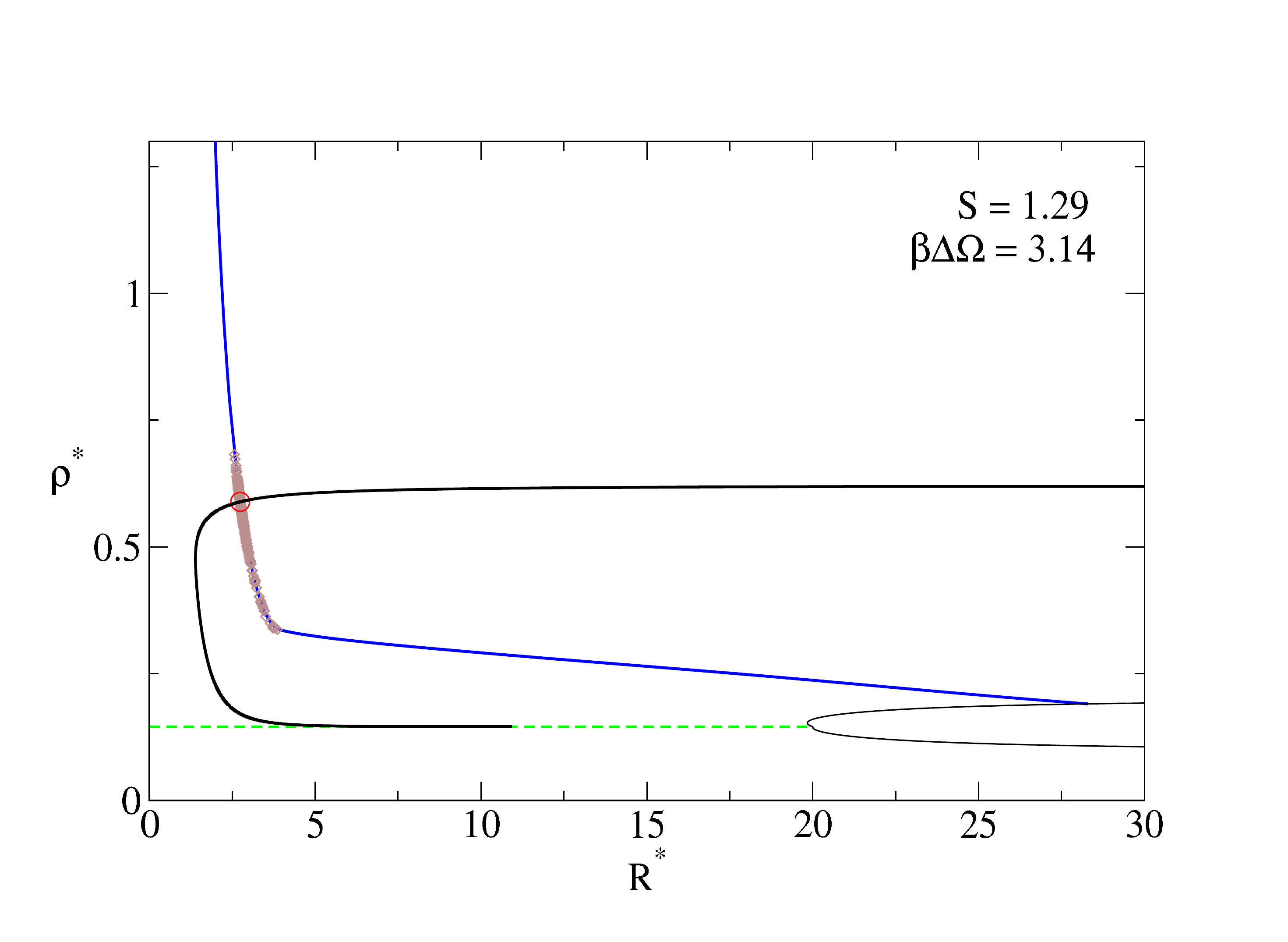}
\includegraphics[angle=0,scale=0.29]{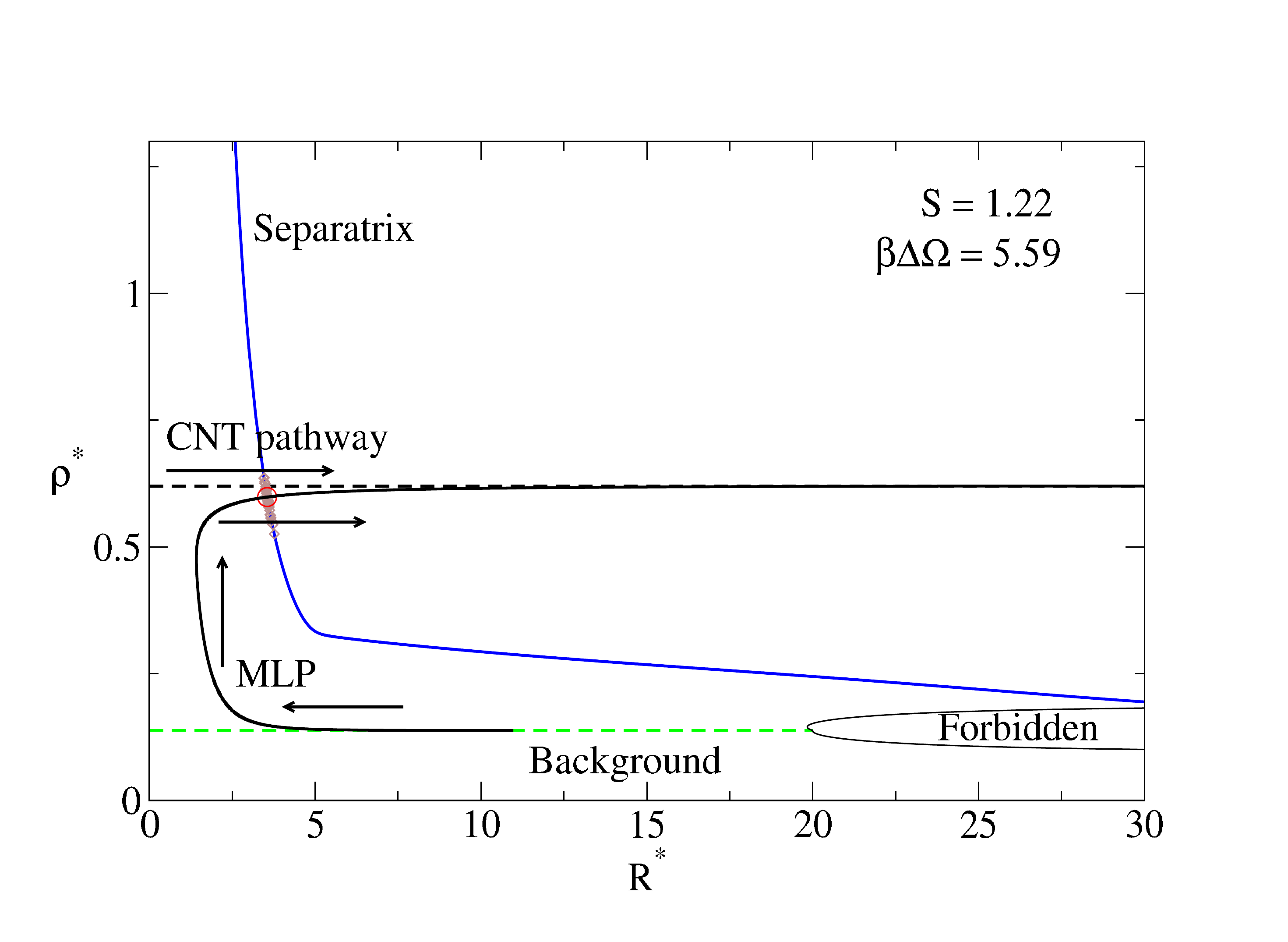}
\caption{The structure of parameter space for the case of nucleation barrier $\beta \Delta \Omega = 3.14$, left panel, and $\beta \Delta \Omega = 5.59$, right panel. The initial state is one of constant density indicated by the lower dashed line. The MLP, as calculated from Eq.(\ref{MLP}), is shown and, in the right hand panel, its three elements are indicated by arrows: the initial compactification stage that begins at large radius and small density deviation from the background, the densification stage and the final growth stage. The latter is very close to the CNT pathway, indicated by the upper dashed line. The separatrix and forbidden regions are shown as is the critical cluster (the circle at the intersection of the separatrix and the MLP). Finally, the figures also show the crossing points from numerous numerical simulations of the stochastic model (open diamonds). For the smaller cluster, on the left, these are widely distributed along the separatrix whereas for the slightly larger cluster on the right, they are already highly concentrated near the critical cluster.}
\label{geometry}
\end{figure}

Figure \ref{geometry} shows examples of the result of this procedure. It is
perhaps surprising that the MLP begins with a cluster having large radius but
very small density change from the background: what might be termed a
long-wavelength density fluctuation. The MLP then passes through three phases.
In the first, the radius decreases while the density increases slowly. In the
second phase, there is a rapid densification with little change in radius. The
third phase involves growth of the cluster with very slow change in density,
which is near that of the dense phase, and is very close to the CNT pathways
(as shown in the Figure). This three-stage process is qualitatively identical
to what is seen using the full fluctuating hydrodynamic
theory\cite{Lutsko_JCP_2011_Com,Lutsko_JCP_2012_1} and so gives us confidence
in the relevance of our model.

We have also performed numerical simulations of the stochastic model (i.e.
Eq.(\ref{SDE})). For these, which we term ``brute force simulations'', the
system began in a random state near the background and was allowed to evolve
until the trajectory crossed the separatrix. Figure \ref{geometry} shows the
crossing points obtained from numerous independent simulations. For the
smallest cluster, with energy barrier $\Delta\Omega\approx2 k_{B}T$, the
crossing points are widely distributed along the separatrix and therefore show
the breakdown of the weak noise assumption. However, even for a slightly
larger barrier, $\Delta\Omega\approx5.5$, the crossings are localized very
near to the critical cluster, providing evidence that the weak-noise
approximation is applicable.

The brute force method is only practical when the energy barriers are
relatively small since the mean first passage time increases exponentially
with the height of the barrier. We therefore have used Forward Flux Sampling
(FFS)\cite{FFS} to follow the behavior of systems with larger barriers. The
procedure begins with a definition of the metastable basin: we take this to be
the region bounded by the $\Delta\Omega= 1 k_{B}T$ free energy line (with
density greater than the background). Next, it is necessary to define a set of
intermediate surfaces (lines in the two-dimensional case) between that
defining the metastable basin and the separatrix. We take advantage of the
similarity in shape between these two lines (compare Figures
\ref{fig_fe_surface} and \ref{geometry}) to construct nested surfaces via a
simple interpolation. Specifically, we pick out $100$ points along each curve
an pair them. We let parameter $u = 0$ correspond to the boundary of the
metastable basin and $u = 1$ to the separatrix: a surface for any intermediate
value of $u$ is constructed by interpolating along the line joining each pair
of points to get an appropriate intermediate point a (Euclidean) distance $u$
between the two end points. These points are used to construct a cubic spline
which defines the intermediate surface. In this way, a sequence of nested
surfaces corresponding to $u =0, u_{1}, ..., u_{n}=1$ are constructed.

Once the surfaces are constructed, we then perform a simulation starting from
a random point within the metastable basin. After an initial ``warmup''
period, during which the initial condition is forgotten, we monitor each time
the stochastic trajectory exits the metastable basin (i.e. crosses the
defining surface in the right direction). This is done until a population of
$N_{0}$ such crossings have been recorded. The total time required for this to
occur, $t_{0}$ is also recorded. We then perform a new series of simulations
in which the initial condition is randomly chosen from this population of exit
points. If the trajectory returns to the metastable basin, the simulation is
terminated. If it first crosses the next surface, then the simulation is again
terminated with the crossing point being recorded. This continues until a
prescribed number, $N_{1}$, of crossings has been observed. Then, the process
is repeated with the crossings of the first surface as the initial condition
and the second surface as the end point, etc., until the separatrix is
reached. Further details are discussed, e.g., in Ref. \cite{FFS}. In this way,
one obtains information both about the nucleation pathways and the mean first
passage time.

\begin{figure}
[ptb]\includegraphics[angle=0,scale=0.15]{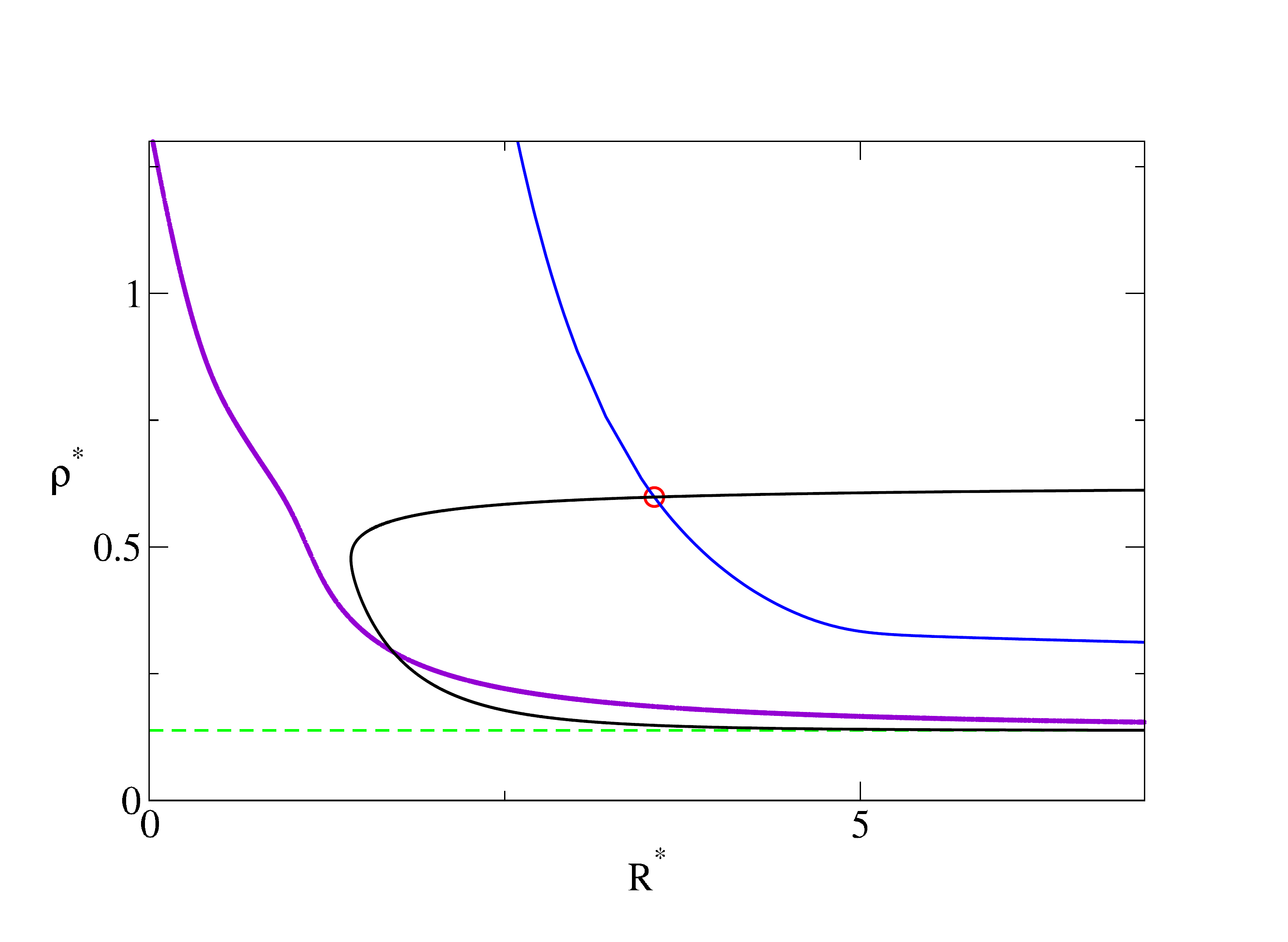}
\includegraphics[angle=0,scale=0.15]{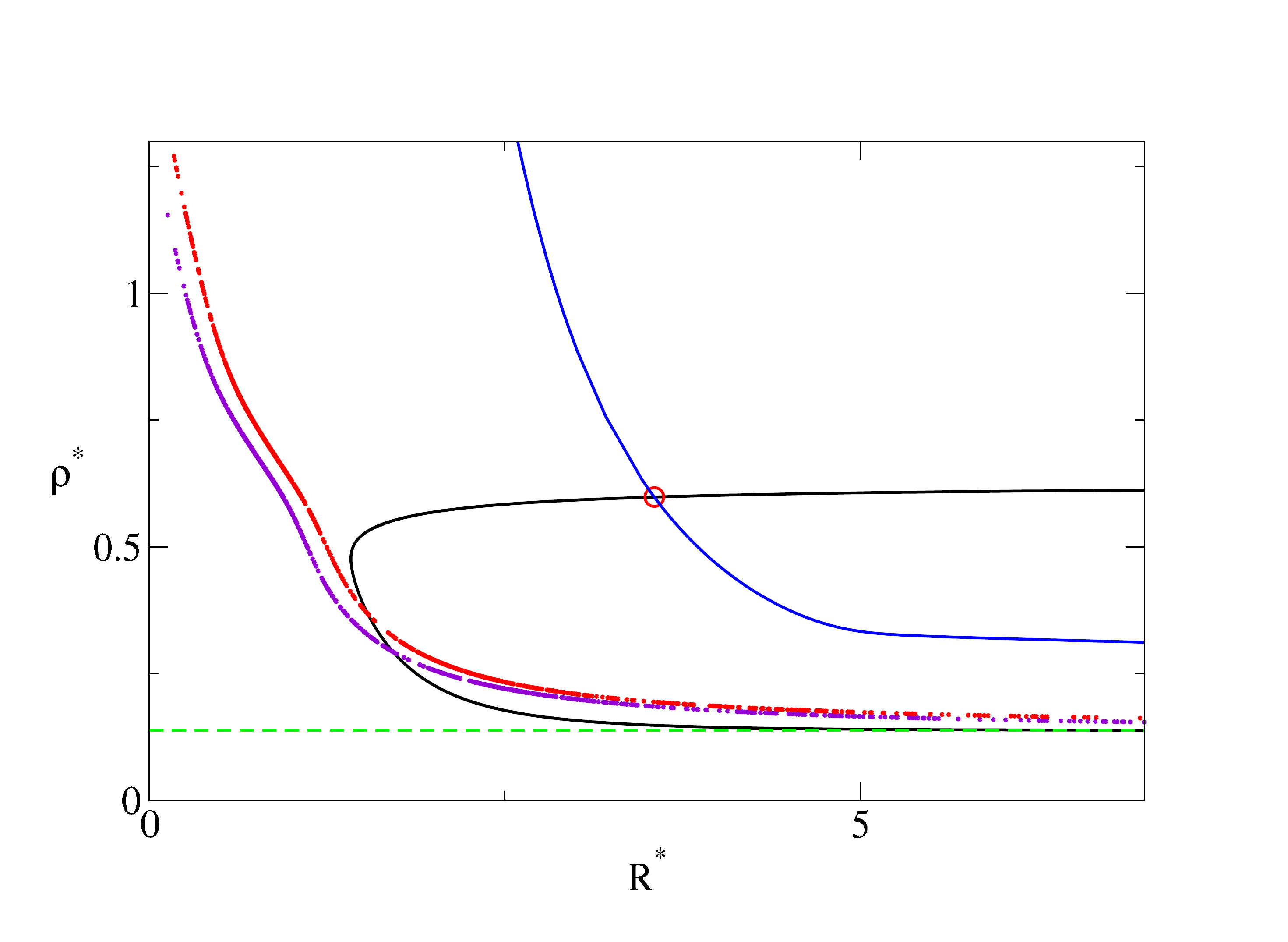}
\includegraphics[angle=0,scale=0.15]{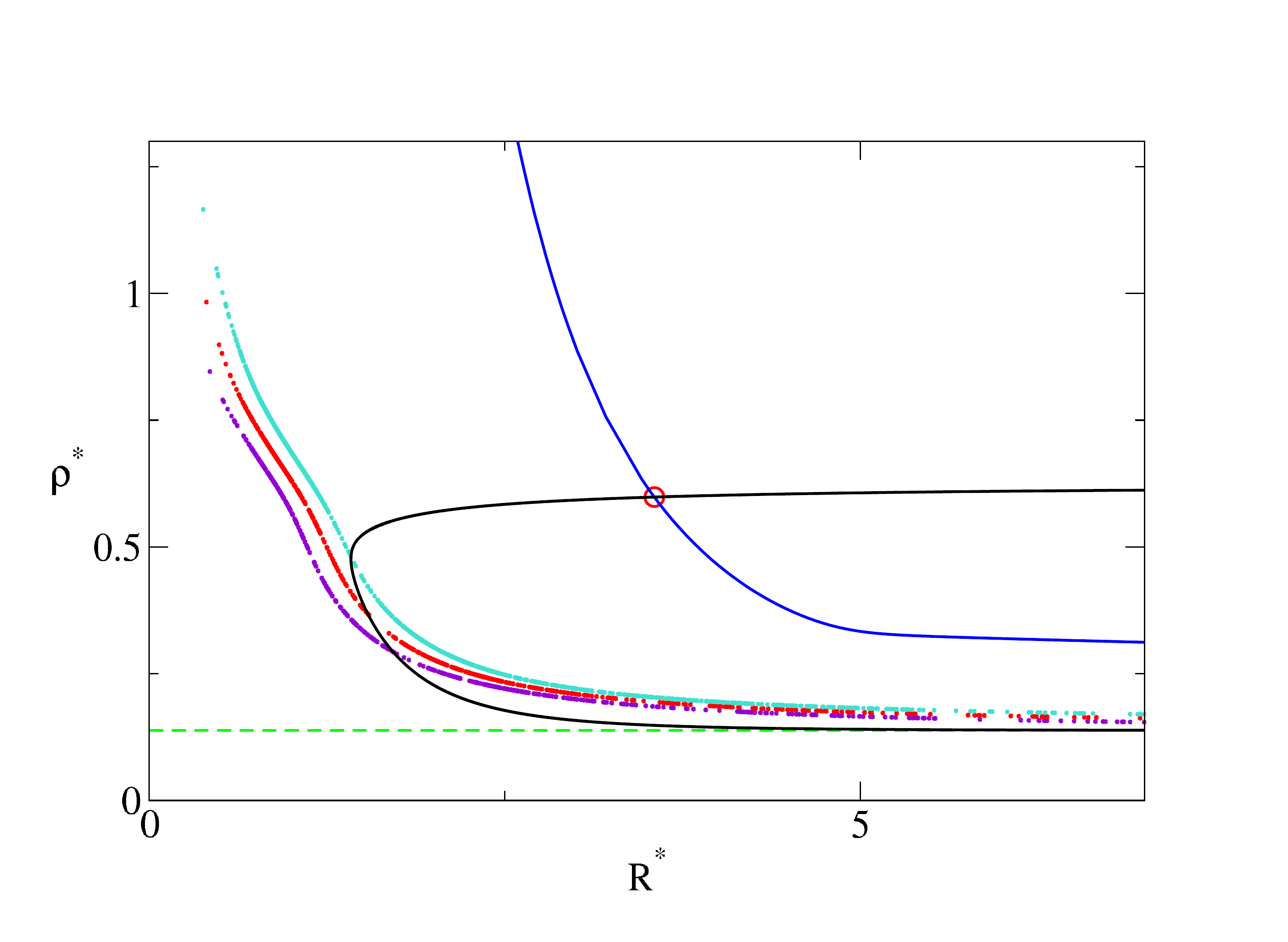}
\caption{Three steps in the FFS process. The leftmost panel shows the points of exit from the metastable basin; the points map out the border of the metastable state. The central panel shows the points of exit from the second surface. One can see that the population of exit points from the metastable basin has been reduced since not all points generate trajectories that reach the second surface. The rightmost panel shows the next step in the process and how it causes removal of points from both of the previous populations.}
\label{fig_FFS_SEQ}
\end{figure}

Figure \ref{fig_FFS_SEQ} shows the way that FFS provides information about the
pathway. First, the distribution along the boundary of the metastable region
is constructed. Then, this population is used as initial conditions in
determining the next set of crossings. Some of the points in the initial
population do not spawn trajectories that reach next surface and so they are
eliminated. During the next round, some of the points on the second surface do
not spawn trajectories crossing the third, so they are dropped. As a result,
some of the points on the first surface no longer have daughter points on the
second surface so they are also dropped. In this way, each stage results in a
refinement that propagates all the way back to the initial population. In the
end, all that are left are points that originated a trajectory that reached
the separatrix.

\begin{figure}
[ptb]\includegraphics[angle=0,scale=0.2]{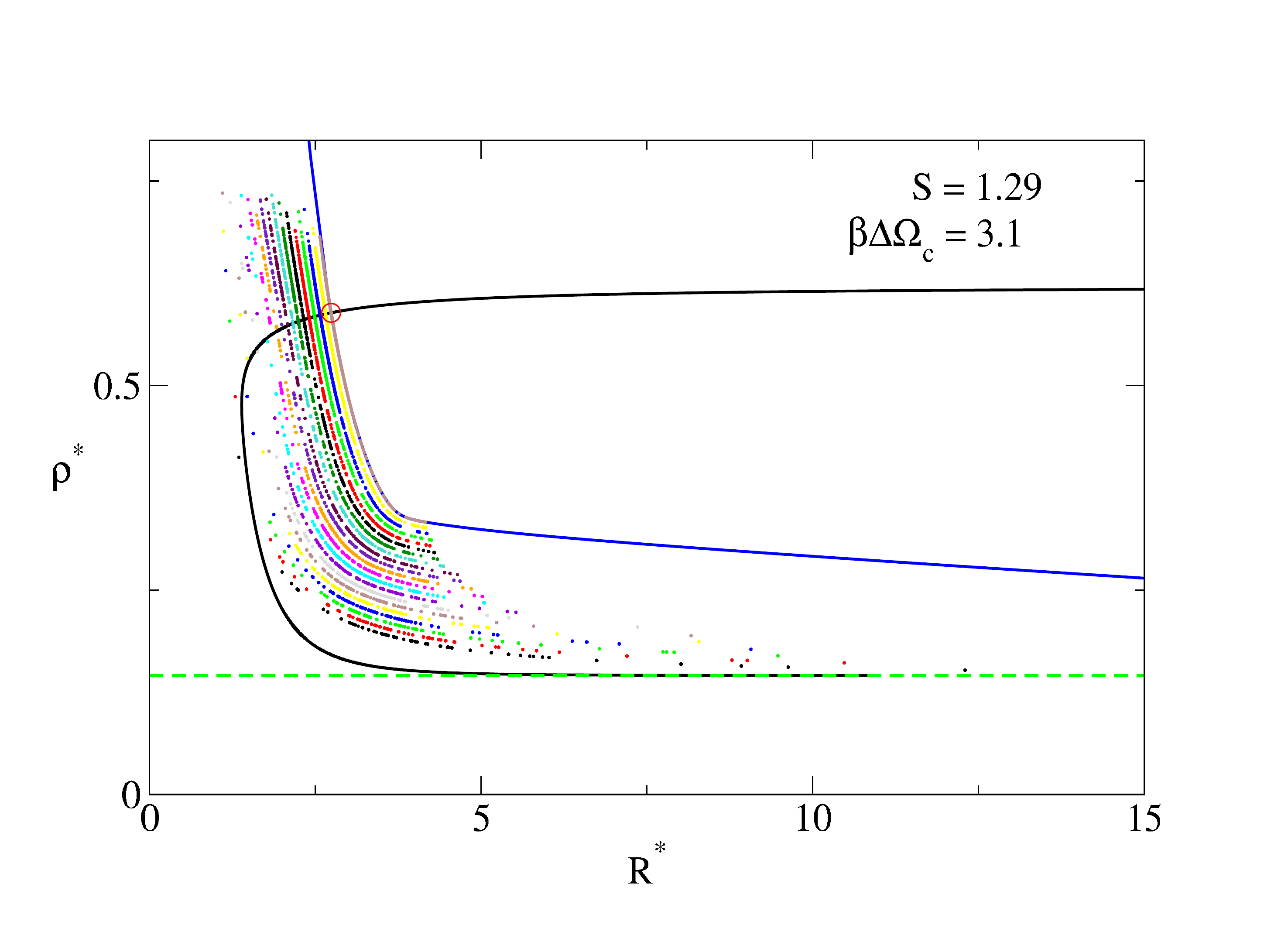}
\includegraphics[angle=0,scale=0.2]{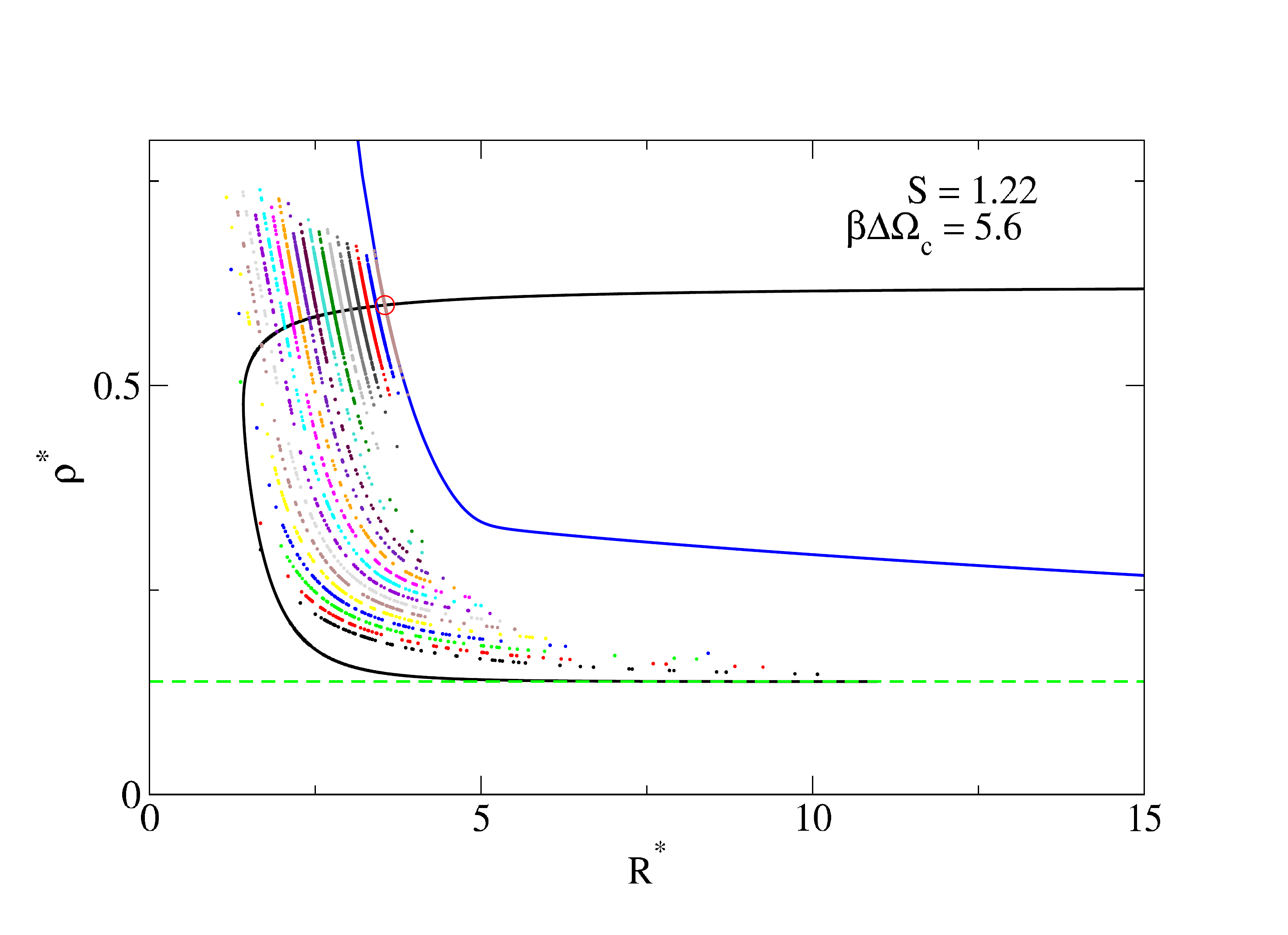}
\includegraphics[angle=0,scale=0.2]{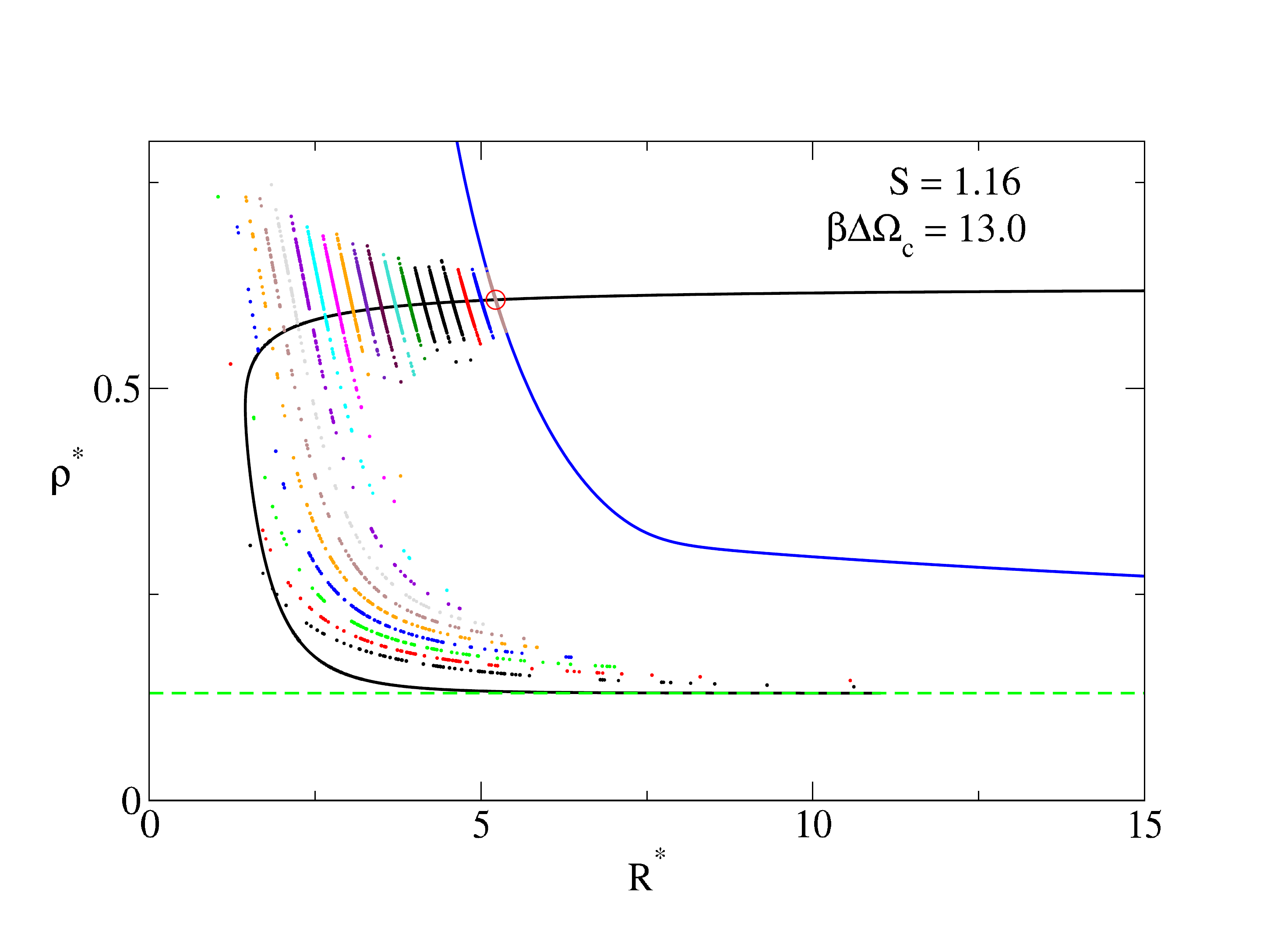}
\includegraphics[angle=0,scale=0.2]{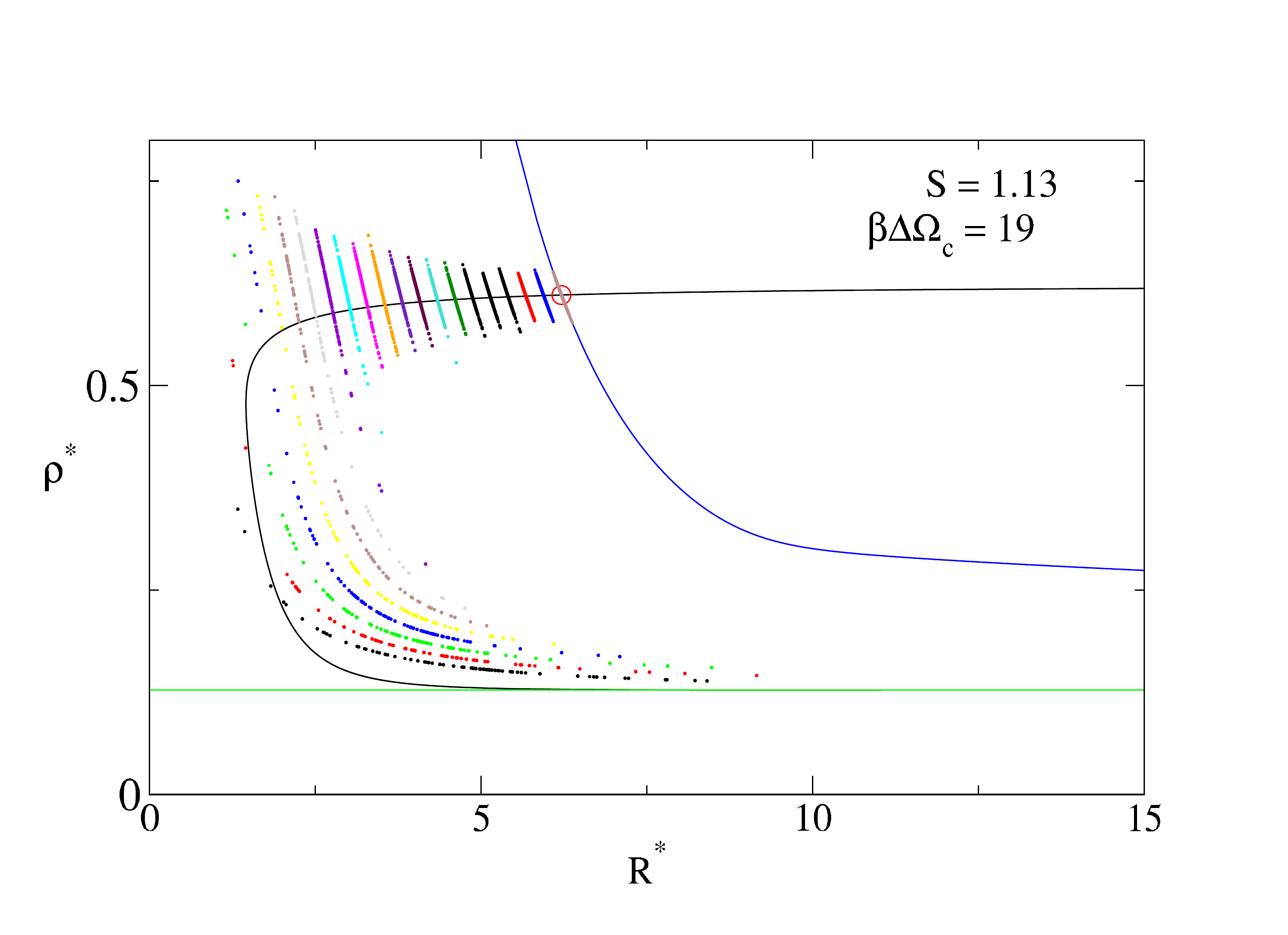}
\caption{The FFS results for the nucleation pathway for several values of supersaturation spanning the range from very low to moderate nucleation barriers. The energy barriers in each case are given on the figures. At high supersaturation, there is little structure to the nucleation pathway and the crossings of the separatrix are highly dispersed; even at slightly higher supersaturations, the pathway becomes more consistent with the predicted MLP and the crossing of the separatrix is localized to the neighborhood of the critical cluster. Note that the spread of exit points on the separatrix at lower supersaturation is consistent with the results of the brute force calculations as shown in Fig. \ref{geometry}. }
\label{fig_FFS}
\end{figure}

Figure \ref{fig_FFS} shows the FFS information for the nucleation pathway for
several different energy barriers. For the most strongly supersaturated
systems, there is little structure apparent except for the fact that nearly
all successful trajectories begin with low density and finite radii: there are
no representative points from the high density, small radius part of the
metastable boundary. At lower supersaturation, the structure becomes more
apparent and is clearly similar to the MLP calculated in the weak noise limit.
The main difference from the MLP is that the densification part of the
transition occurs at a somewhat larger radius than the weak noise theory predicts.

\begin{figure}
[ptb]\includegraphics[angle=0,scale=0.4]{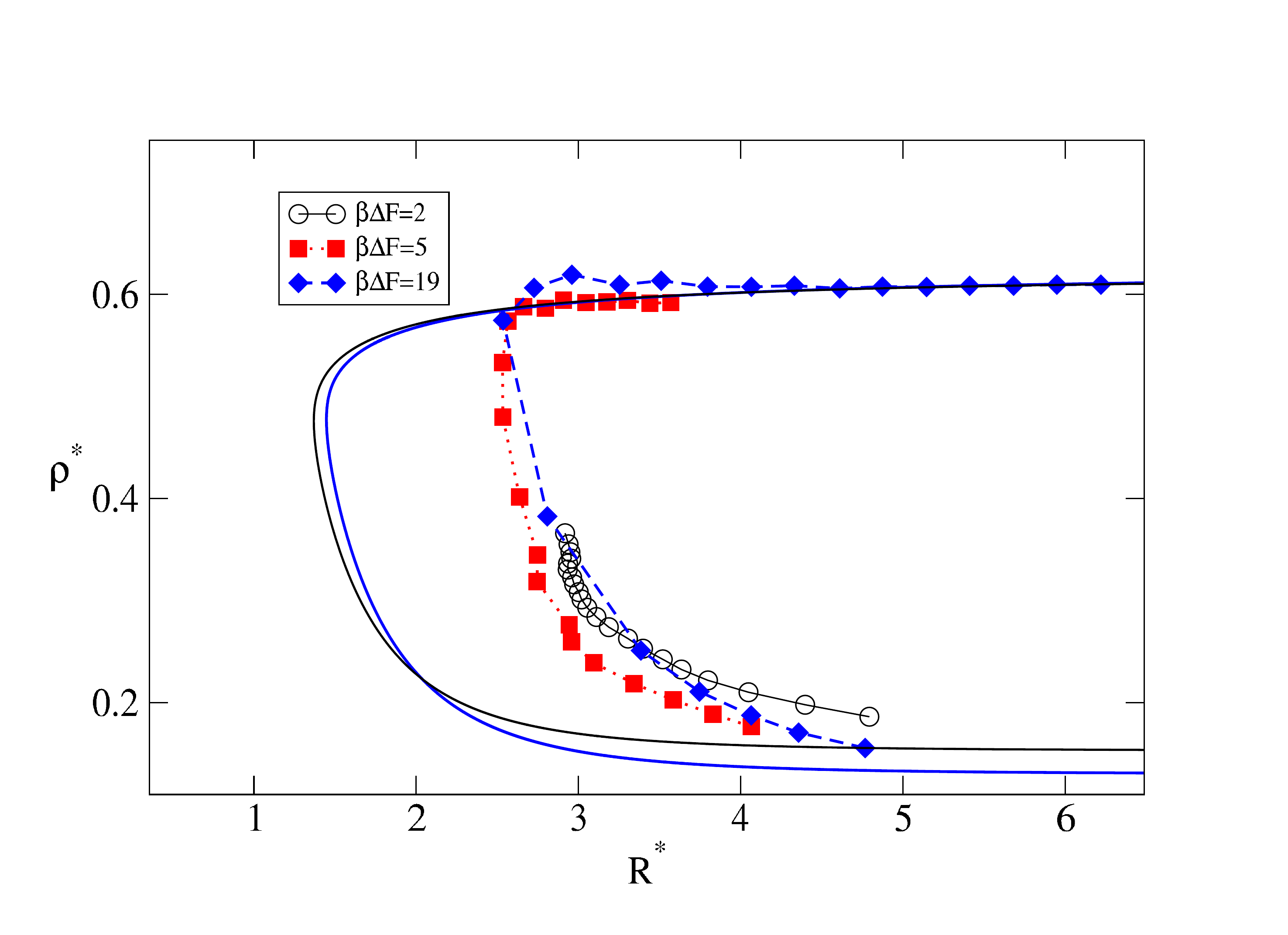}
\caption{The average pathways as determined from the Forward Flux Sampling simulation data for several values of the nucleation barrier (symbols connected with lines) at $T^{*} = 0.40$ . The calculated most likely paths for $\beta \Delta \Omega = 2,19$ are shown as well (full lines). }
\label{averagepaths}
\end{figure}

\begin{figure}
[ptb]\includegraphics[angle=0,scale=0.4]{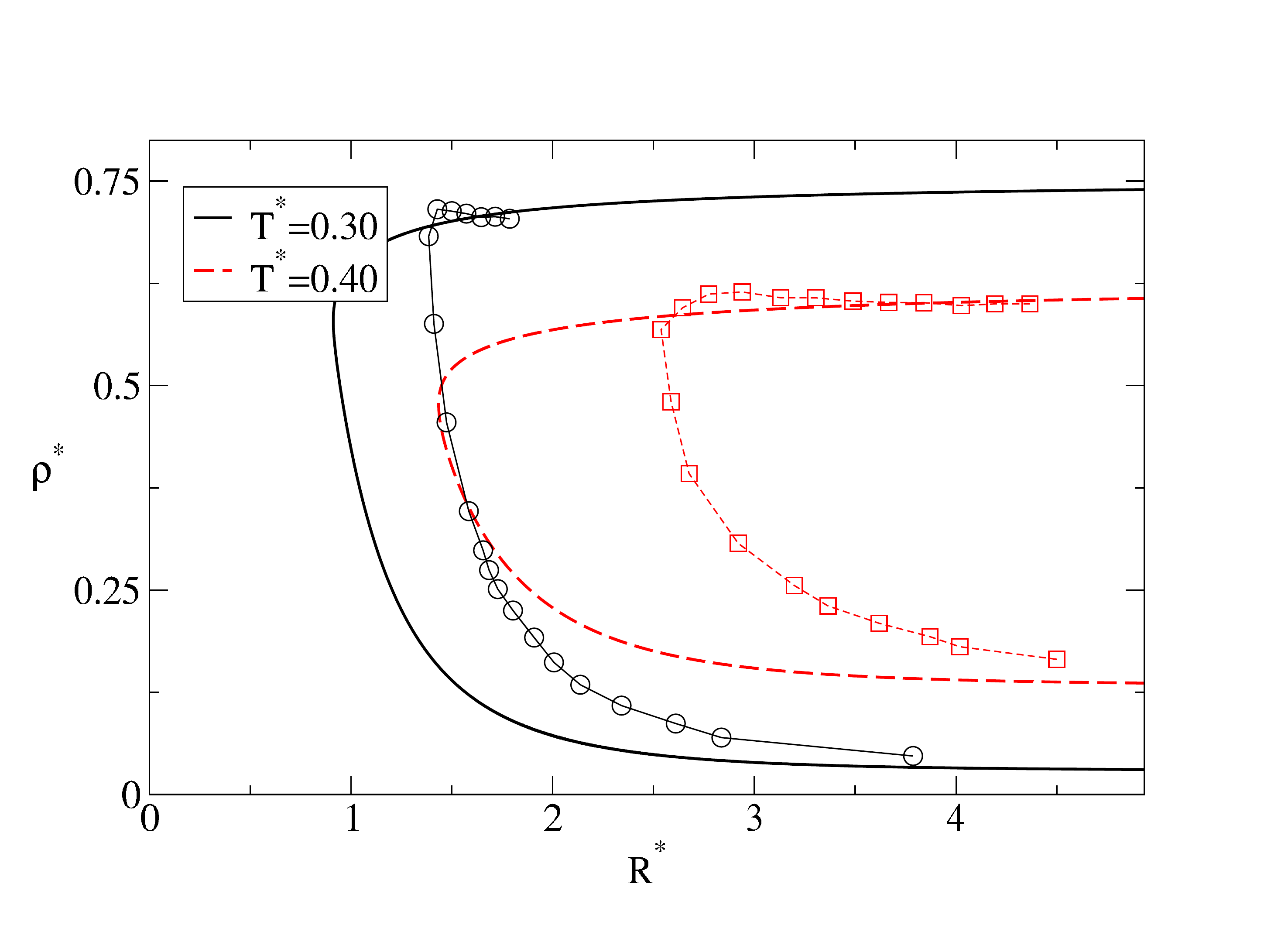}
\caption{The average pathways as determined from the simulation data for several values of a nucleation barrier of $\Delta \beta F = 8.2$ at two different temperatures, $T^{*} = 0.30,0.40$. The full lines are the predictions based on the weak-noise approximation and the symbols show the simulation results. Convergence of the simulations to the weak-noise limit as the temperature is lowered is evident.}
\label{AveragePathsConverge}
\end{figure}

A comparison of the calculated MLP and the \textit{average pathways} as
determined from the FFS simulations using a method discussed in Appendix
\ref{AveragePaths} is shown in Fig.\ref{averagepaths}. The pathways calculated
in the weak-noise limit are in semi-quantitative agreement with the observed
pathways although the latter show a systematic shift to larger radii. It is
possible that this difference is due to the average pathway being distinct
from the most likely pathway but it seems consistent with the raw data as
shown in Fig. \ref{fig_FFS} above. In principle, we could determine the most
likely pathway but in attempting to do so we found that our data were too
noisy to give convincing results. Figure \ref{AveragePathsConverge} shows the
average paths for a nucleation barrier of $\Delta F = 8.2 k_{B}T$ for
temperatures $T^{*} = 0.40$ and $0.30$. At the lower temperature, the MLP
shifts towards smaller radii by about $0.5\sigma$ while the shift in the
average path is about twice this, thus verifying the expected convergence of
the average path to the weak noise result as the temperature is lowered.

\subsection{Nucleation Rate}

The most important quantity, for practical applications, is the nucleation
rate. Here, we compare the time required for nucleation in three
approximations. The first is the standard version of CNT for the case of
diffusion limited nucleation. In this case, the exact mean first passage time
can be calculated numerically using Eq.(\ref{mfpt_exact}). The second
approximation is a modified version of CNT that uses the same model for a
mass-conserving cluster as is used in the two-variable theory but with the
interior density fixed at the bulk value (as in the usual CNT). This simply
means that we replace the free energy and the value of $g_{RR}$ used to
reproduce CNT by the equivalent functions evaluated using the mass-conserving
cluster profile as inputs to evaluate Eq.(\ref{mfpt_exact}). The third
approximation is the two-variable model for which we only have the approximate
expression for the mean first passage time (MFPT), Eq.(\ref{T}).

\begin{figure}
[ptb]\includegraphics[angle=0,scale=0.3]{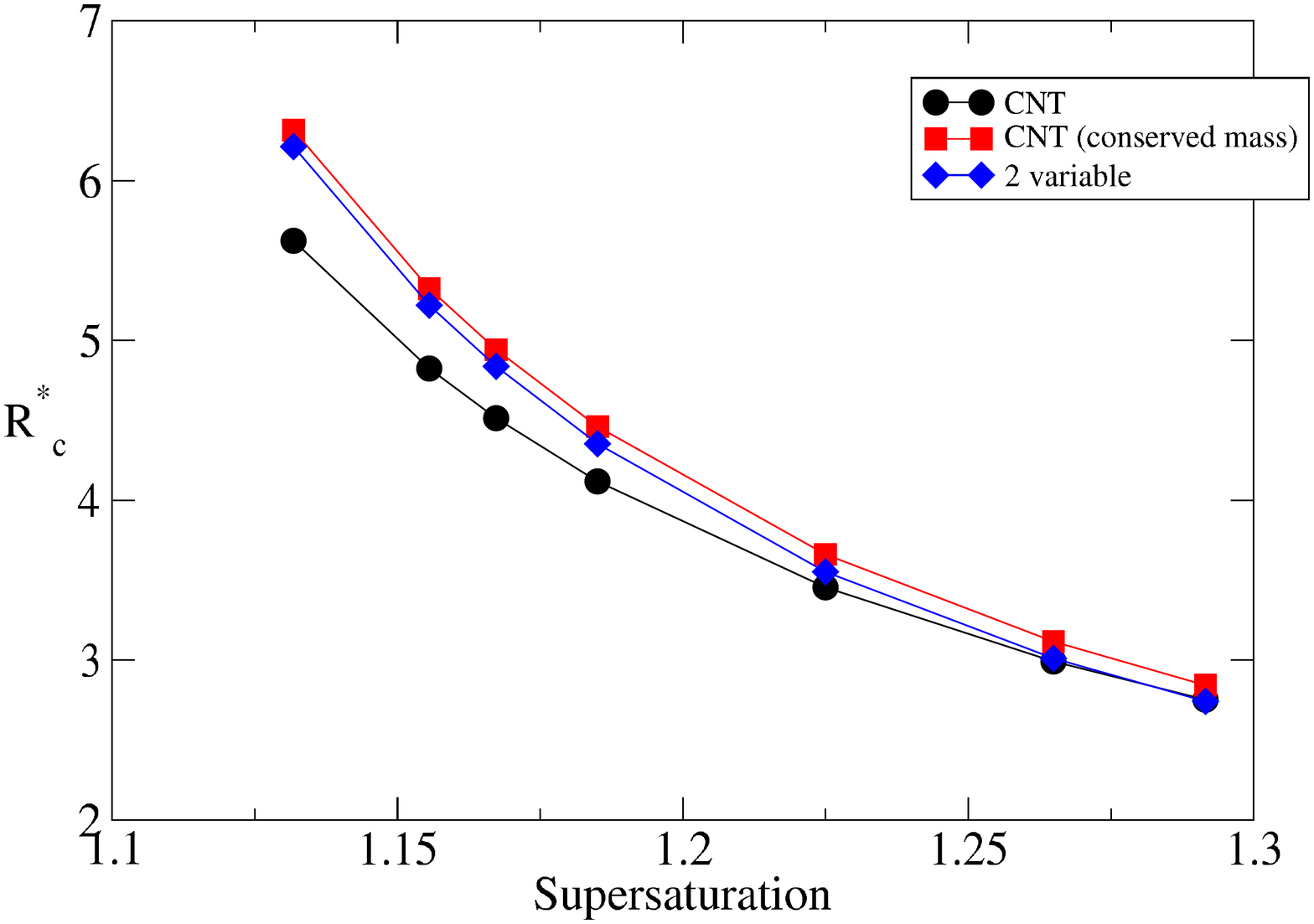}
\includegraphics[angle=0,scale=0.3]{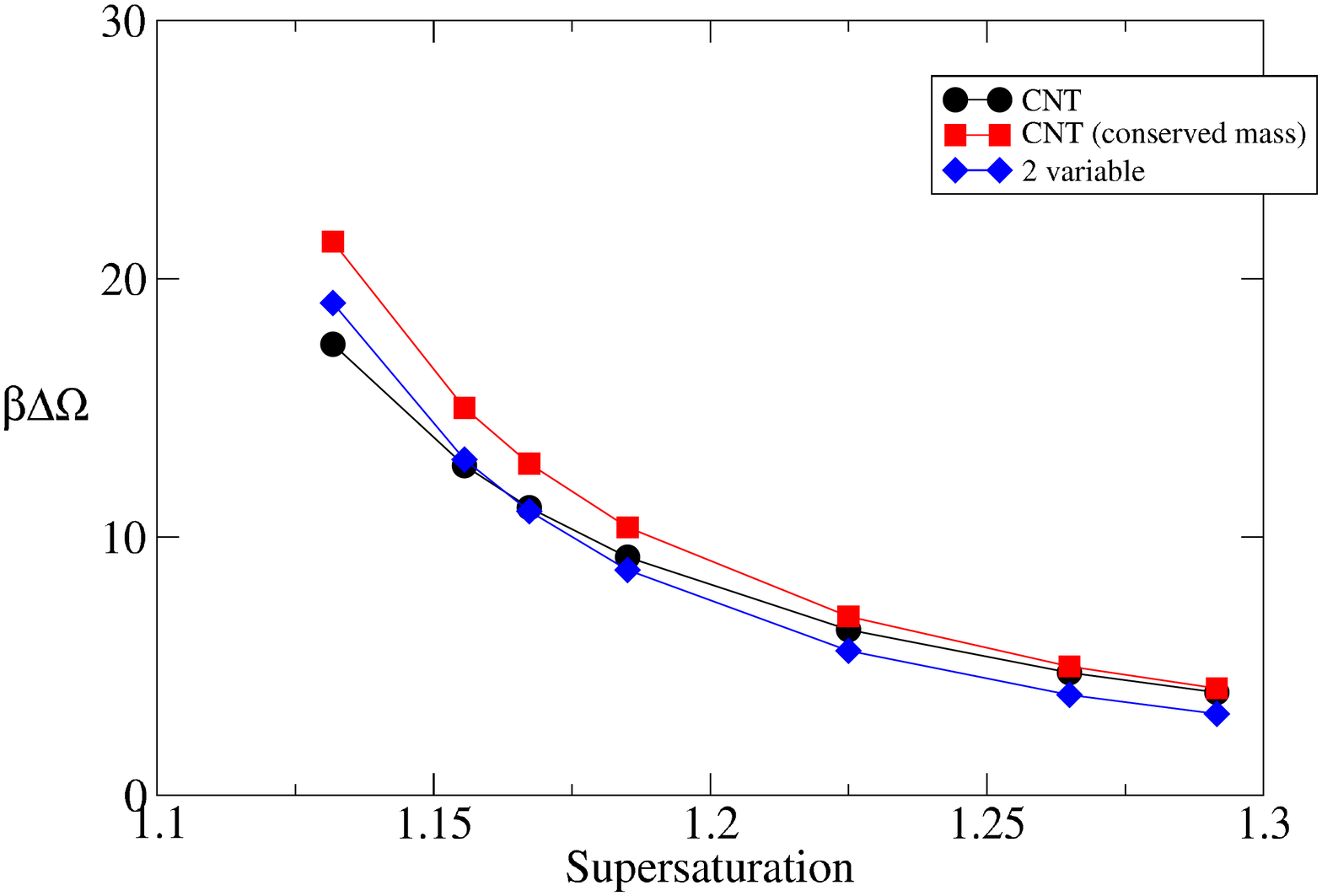}
\caption{The critical radii (left panel) and energy barriers (right panel) as functions of the supersaturation  $S \equiv \rho/\rho_{coex}$. Results are shown are the usual CNT values, Eqs.(\ref{Rcnt},\ref{Ecnt}) determined by setting $\frac{d}{d R}\beta \Omega = 0$ with the density fixed at that of the bulk liquid, the same calculation but using the mass-conserving form of the free energy, Eq.(\ref{Emassconserving}), and the result of setting to zero variations of  the mass-conserving free energy with respect to both radius and interior density.}
\label{barrier}
\end{figure}

Figure \ref{barrier} shows the radius and excess free energy of the critical
cluster as functions of the supersaturation in these three approximations.
There is little difference in the radius in the three approximations. The
difference in the excess free energy is very small for the highest
supersaturations and amounts to about $4k_{B}T$ for the lowest
supersaturation. The excess free energy for the two-variable theory is always
less than for the mass-conserving CNT as would be expected due to the
additional degree of freedom. On the other hand, the usual (non-mass
conserving) CNT energy barrier is sometimes higher than for the two-variable
theory but not always. This is because with mass-conservation, there is an
additional free-energy penalty when forming a cluster due to the cost of the
depletion zone.

\begin{figure}
[ptb]\includegraphics[angle=0,scale=0.4]{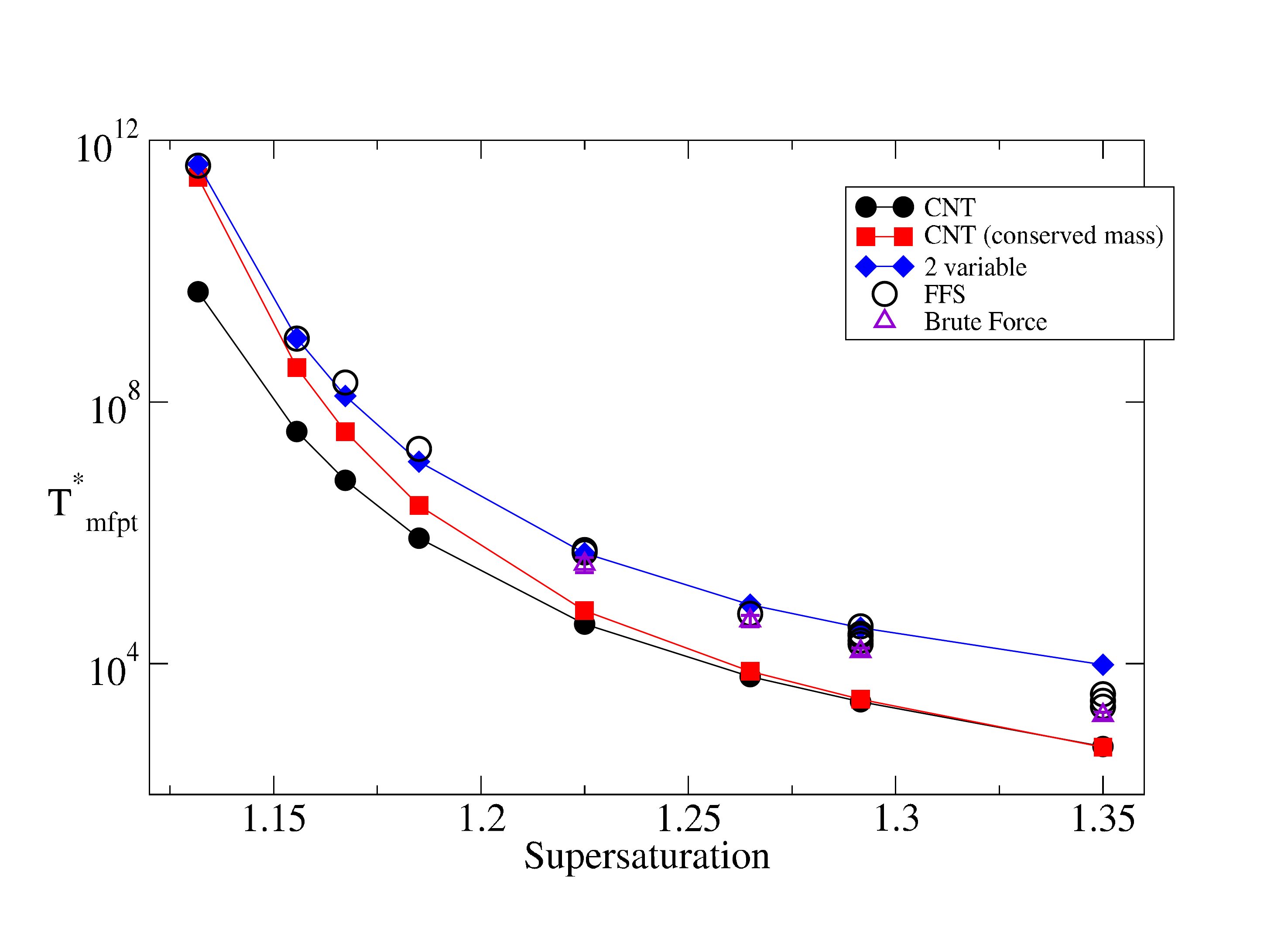}
\caption{The mean first passage time as a function of the supersaturation  $S \equiv \rho/\rho_{coex}$ as determined from the usual CNT, Eq.(\ref{mfpt_exact}), CNT with the mass-conserving form of the free energy, Eq.(\ref{Emassconserving}), the two-variable theory, Eq.(\ref{T}), and the results of FFS and brute force simulations of the two-variable theory. The brute force values are the result of averaging over an ensemble of approximations 100 trajectories and the standard-deviation is on the order of the symbol-size. At the higher values of supersaturation, multiple FFS determinations were performed varying the random-number generator seed, the number of energy surfaces and the number of crossings of each surface thus giving multiple FFS symbols for some supersaturations: the spread of these results gives some idea of the accuracy of our FFS values.}
\label{mfpt}
\end{figure}

\begin{figure}
[ptb]\includegraphics[angle=0,scale=0.4]{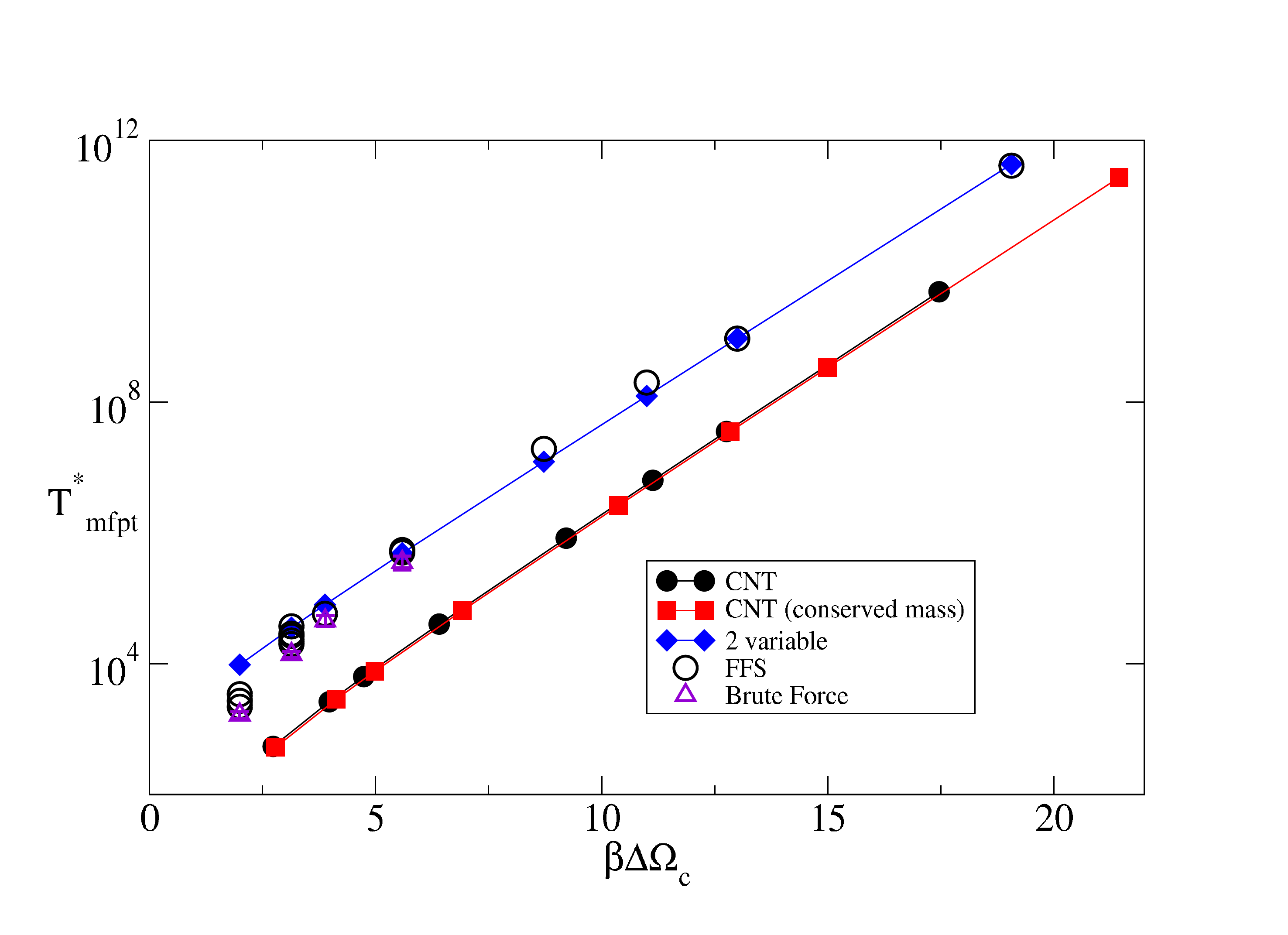}
\caption{The same as Fig. \ref{mfpt} but showing the mean first passage time as a function of the nucleation barrier. There is a systematic difference between the one- and two-variable theories that cannot be attributed solely to differences in the free energy of the critical cluster.}
\label{emfpt}
\end{figure}

Figure \ref{mfpt} shows the MFPT as a function of supersaturation as
determined from the three approximations as well as from simulation using FFS
and, for the systems with the smallest barriers, brute force. First, we note
that the FFS and brute force values are mutually consistent thus providing
evidence that the FFS implementation is accurate. Second, the times determined
from FFS are in remarkable agreement with the approximate times calculated for
the two-variable theory except, perhaps, at the highest supersaturation. Given
that the analytic theory is only valid in the weak noise limit, this agreement
is surprisingly good. Fourth, the mass-conserving CNT seems to converge
towards the two-variable theory as supersaturation decreases (i.e. as the
critical cluster grows) in line with the general expectation that CNT becomes
exact in the limit of large critical clusters. Finally, the MFPT given by the
usual CNT is consistently between one and two orders of magnitude below the
actual value. This type of discrepancy between CNT and experiment or
simulation have often been noted in the literature (some examples are Ref.
\onlinecite{PhysRevE.56.5615,Tanaka2011,Tanaka2014}) and are typically
attributed to inaccuracies in the calculation of the free energy barrier.
However, we show in Fig. \ref{emfpt} the MFPT as a function of the barrier
used in the various approximations and it can be seen that this accounts for
the difference between CNT with and without mass-conservation, but that in
fact the difference from the 2 variable theory cannot be explained on this
basis since even with the same energy barrier, the CNT results are
consistently between one and two orders of magnitude below the actual MFPT. We
conclude that the difference is mostly attributable to the different
\textit{dynamics} and that the inaccuracies in the calculation of the free
energy are of secondary importance.

\section{Discussion of the results}

In this Section, we summarize and attempt to rationalize the various results
obtained from our calculations.

\subsection{Properties of the critical cluster}

\begin{itemize}
\item The radius of the critical cluster predicted by CNT is somewhat smaller
than that predicted by the mass-conserving theories while there is little
difference between the mass-conserving one- and two-variable theories. This
larger critical radius is due to the need to offset the increased free-energy
penalty arising from the depletion zone.

\item The CNT free energy barrier is always smaller than that of the
mass-conserving one-variable theory since the latter has an additional free
energy penalty due to the lower vapor density in the depletion region. The two
variable mass conserving theory has an additional degree of freedom relative
to the one-variable mass-conserving theory, so it always has a lower free
energy barrier. In the comparison between CNT and the two-variable theory, the
two effects (greater free energy penalty and more degrees of freedom) are in
play and there is no definitive trend: for smaller clusters, the two-variable
theory has lower barrier while for larger clusters, CNT has the smaller
barrier. This competition and potential canceling of the two effects can help
explain why CNT sometimes appears to be quite accurate.
\end{itemize}

\subsection{Mean first passage times}

\begin{itemize}
\item Figure \ref{mfpt} shows that the mass-conserving one-variable theory
always has larger MFPT (lower nucleation rate) than does CNT and Fig.
\ref{emfpt} shows that this is entirely attributable to the difference in free
energy barrier. This is therefore an example where ``correcting'' the
calculation of the free energy barrier and using the CNT rate formulas gives
the correct nucleation rate.

\item The same figures show that the differences between the one-variable and
two-variable theories are systematic and cannot be explained solely in terms
of poor estimation of free energies. These differences must be attributable to
the combination of (a) the calculation of the flux through the critical
cluster and (b) the description of the metastable state.

\item The calculation of the flux through the critical cluster cannot explain
the differences at lower supersaturation (larger nucleation barrier and
critical cluster radius). This is because the most likely path, as calculated
in the weak noise approximation, pass through the critical cluster in the
direction of of the most unstable eigenvector and the figures show that this
is nearly parallel to the $R$-axis. In other words, the eigenvalue that occurs
in the formula for the MFPT is effectively $g^{RR}\frac{\partial^{2} \Delta
F}{\partial R^{2}}$ which is the same quantity as is used in the evaluation of
the CNT nucleation rate.

\item Consequently, the differences in nucleation rates must be attributable
to the calculation of population in the metastable basin. This is clearly very
different in the two cases (an exponential distribution in $R$ for CNT versus
an algebraic distribution in $R$ for the two-variable theory). Ultimately,
this is a reflection of the fact that a theory with long-wavelength, low
free-energy fluctuations admits of a much larger accessible phase space than
does one with only an isolated minimum on a one-dimensional curve. Put
differently, it highlights the failure, within this model, of the common
assumption that the probability to observe a cluster of size $N$ is, up to
normalization, $e^{-\beta F(N)}$.

\item The apparent convergence of the one-variable and two-variable results in
Fig. \ref{mfpt} appears to be accidental: the systematic over-estimation of
the nucleation barrier in the one-variable theory compensates for the
under-estimate of the population of the metastable basin.
\end{itemize}

\subsection{General comments concerning the pathways}

\begin{itemize}
\item The most likely path calculated in the weak noise limit gives a good
semi-quantitative description of the observed average nucleation pathway. Both
show the three step process of long-wavelength density fluctuation that
contracts followed by densificiation followed by the CNT pathway.

\item Both the calculated MLP and the observed average pathways show a minimum
cluster size which is consistent with the observations of Trudu et al.
\cite{Trudu} in simulations of crystallization.

\item Since the density fluctuations that become clusters begin with very small density, the usual criterion used in computer simulations for identifying a cluster - namely a group of molecules with local density above some threshhold - is not a good order parameter. 

\item The numerical results show that the both the calculated most-likely
nucleation pathway and the observed average pathway are virtually insensitive
to the supersaturation.

\item The effect of varying the supersaturation can be viewed as simply moving
the position critical cluster along the pathway.

\item Only for very small clusters, with barrier less than $5 k_{B}T$ do we
see any significant crossing away from the neighborhood of the critical cluster.
\end{itemize}

\section{Conclusions}

In this paper we have developed a two-variable description of
diffusion-limited nucleation based on a formalism developed using fluctuating
hydrodynamics. From a purely theoretical perspective, we noted that the most
naive generalization of CNT is unphysical. We suggested that this was due to
neglect of mass conservation and developed a mass-conserving model that gives
qualitatively similar results to those obtained from a direct analysis of the
underlying fluctuating hydrodynamics. While in some sense a minimal extension
of Classical Nucleation Theory, the model nevertheless illustrates the full
complexity of nucleation including strong noise effects and a continuum of
possible nucleation pathways. We showed by means of comparison between theory
and numerical simulations that both the typical nucleation pathway and the
nucleation rate could be accurately determined using results from the
weak-noise limit. Finally, this model showed similar deviations from CNT as
are commonly reported based on large-scale computer simulation and
experiments. Our main practical conclusion was that these differences could
not be attributed to errors in the calculation of the free energy barrier but,
rather, are a direct result of the more complex dynamics. For this reason, we
believe that nucleation can only be properly understood when both
thermodynamics and dynamics are incorporated in a self-consistent manner.

The model discussed here therefore provides a sort of ``laboratory'' within
which a variety of post-CNT effects can be studied. We also note that, despite
its very different motivations, our model of a cluster - consisting of both a
dense core and a low density depletion zone - shares some qualitative
similarity to the Extended Modified Liquid Drop Model\cite{Reguera2004c} of
Reiss, Reguera and co-workers. The fact that the core density is allowed to
vary and that mass is conserved means that there are also obvious similarities
to the Generalized Gibbs
approach\cite{Schmelzer1987,Schmelzer2006a,Schmelzer2011}. The main
contribution of the present work is to supplement these ideas with a
consistent dynamics that allows for the study of the entire process of
nucleation from the initial density fluctuations to the critical cluster and
beyond. Nevertheless, in order to tame the divergence of the normalization of
the distribution function in subcritical systems while allowing for arbitrary
post-critical cluster growth in supersaturated systems, we were forced to
introduced a somewhat contrived model. Further studies (not reported here)
have shown that the choice of our parameter $\lambda$ has little effect on the
properties of nucleation provided the available mass is sufficient to allow
for the creation of a critical cluster: this is true even for the extreme case
of $\lambda= 0$ which corresponds to a fixed volume. This, together with the
qualitative agreement with similar work based on the full hydrodynamic model,
give us some hope that our results are sufficiently ``generic'' as to be
informative concerning diffusion-limited nucleation in real systems.

\begin{acknowledgments}
The work of JFL was supported in part by the European Space Agency under
contract number ESA AO-2004-070 and by FNRS Belgium under contract C-Net
NR/FVH 972. MD acknowledges support from the Spanish Ministry of Science and
Innovation (MICINN), FPI grant BES-2010-038422 (project AYA2009-10655).
\end{acknowledgments}

\bibliographystyle{plain}
\bibliography{cnt_2var}

\appendix

\section{Density dependence of surface tension}

\label{SurfaceTension} We can gain insight into how the surface tension term
should depend on density by starting from a more microscopic perspective. A
suitable approximation is the squared-gradient free energy,
\begin{equation}
\Omega[\rho] = \int\left(  \omega(\rho(\mathbf{(}r)) + g \left(
\mathbf{\nabla} \rho(\mathbf{r}\right)  ^{2} \right)  d\mathbf{r}.
\end{equation}
This can be derived from exact Density Functional Theory under the assumption
that the spatial variation of the density is sufficiently
slow\cite{Lutsko2011a}. Let us consider the case of a planar interface so that
the density depends only on one Cartesian coordinate, say $\rho(z)$ and
suppose that the density at $z \rightarrow-\infty$ is equal to the equilibrium
density called $\rho_{\infty}$ in the main text. The density at $z
\rightarrow\infty$ will be taken to be some small deviation from this value,
$\rho_{\infty} + \Delta\rho$. Then, The free energy can be expanded up to
second order in the density as
\begin{equation}
\frac{\Omega[\rho]-\Omega(\rho_{\infty})}{A} = \int_{-\infty}^{\infty} \left(
\omega^{\prime\prime}(\rho_{\infty})(\Delta\rho(z))^{2} + g \left(
\mathbf{\nabla} \Delta\rho(z)\right)  ^{2} \right)  dz
\end{equation}
where $A$ is the planar area. The left hand side is just the excess free
energy per unit surface (relative to the background). If we assume that the
transition from densities close to $\rho_{\infty} + \Delta\rho$ is monotonic
and occurs over a spatial region of length $w$, then this will be
\begin{equation}
\frac{\Omega[\rho]-\Omega(\rho_{\infty})}{A} = a_{1} w \omega^{\prime\prime
}(\rho_{\infty})(\Delta\rho)^{2} +a_{2}\frac{1}{w} g ( \Delta\rho)^{2}%
\end{equation}
where $a_{1},a_{2}$ are constants dependent on the precise shape of the
function $\Delta\rho(z)$. This shows that for fixed width, the surface tension
is proportional to the square of the density difference. Rather than fixed
width, we might prefer to fix the width by minimizing the free energy. This
gives
\begin{equation}
w = \frac{a_{2} g }{a_{1} \omega^{\prime\prime}(\rho_{\infty})}%
\end{equation}
which is independent of density, so the conclusion remains the same. Similar
arguments lead to the same conclusion for liquid-vapor interfaces near
coexistence\cite{lutsko:acp,Lutsko2011a}. These facts therefore strongly
suggest that the surface tension should vary as the square of the density difference.

\section{The kinetic coefficients}

\label{Metric} For the density parameterization,
\begin{equation}
\rho\left(  r\right)  =\rho_{0}\Theta\left(  R-r\right)  +\rho_{1}%
\Theta\left(  r-R\right)  \Theta\left(  R_{1}-r\right)  +\rho_{\infty}%
\Theta\left(  r-R_{1}\right)  .
\end{equation}
the cumulative mass is
\begin{align}
\Delta m\left(  r\right)   &  =\left(  \rho_{0}-\rho_{\infty}\right)  \left(
V\left(  r\right)  \Theta\left(  R-r\right)  +V\left(  R\right)  \frac
{R_{1}^{3}-r^{3}}{R_{1}^{3}-R^{3}}\Theta\left(  r-R\right)  \Theta\left(
R_{1}-r\right)  \right)
\end{align}
and
\begin{align}
\frac{\partial\Delta m\left(  r\right)  }{\partial R}  &  =4\pi\left(
\rho_{0}-\rho_{\infty}\right)  \frac{R^{2}R_{1}^{2}}{\left(  R_{1}^{3}%
-R^{3}\right)  ^{2}}\left(  R_{1}\left(  R_{1}^{3}-r^{3}\right)
-\frac{\partial R_{1}}{\partial R}R\left(  R^{3}-r^{3}\right)  \right)
\Theta\left(  r-R\right)  \Theta\left(  R_{1}-r\right) \\
\frac{\partial\Delta m\left(  r\right)  }{\partial\rho_{0}}  &  =V\left(
r\right)  \Theta\left(  R-r\right) \nonumber\\
&  +V\left(  R\right)  \frac{1}{R_{1}^{3}-R^{3}}\left(  \left(  R_{1}%
^{3}-r^{3}\right)  -\frac{\left(  R^{3}-r^{3}\right)  }{\left(  R_{1}%
^{3}-R^{3}\right)  }3R_{1}^{2}\left(  \rho_{0}-\rho_{\infty}\right)
\frac{dR_{1}}{d\rho_{0}}\right)  \Theta\left(  r-R\right)  \Theta\left(
R_{1}-r\right) \nonumber
\end{align}
A straightforward calculation then gives
\begin{align}
\label{gbig}g_{RR}  &  =\frac{4\pi}{5}\frac{\left(  \rho_{0}-\rho_{\infty
}\right)  ^{2}}{\left(  \rho_{\infty}-\rho_{0}y^{3}\right)  }R^{3}\frac
{1}{\left(  1+y+y^{2}\right)  ^{3}}\left(
\begin{array}
[c]{c}%
\left(  5+6y+3y^{2}+y^{3}\right)  +3y^{2}\left(  1+3y+y^{2}\right)
\frac{dR_{1}}{dR}\\
+y^{3}\left(  1+3y+6y^{2}+5y^{3}\right)  \left(  \frac{dR_{1}}{dR}\right)
^{2}%
\end{array}
\right)  \allowbreak\allowbreak\\
g_{R\rho_{0}}  &  =\frac{4\pi}{30}\frac{\left(  \rho_{0}-\rho_{\infty}\right)
}{\rho_{\infty}-\rho_{0}y^{3}}R^{4}\frac{1-y}{\left(  1+y+y^{2}\right)  ^{2}%
}\left(
\begin{array}
[c]{c}%
2\left(  5+6y+3y^{2}+y^{3}\right) \\
+3y^{2}\left(  1+3y+y^{2}\right)  \frac{dR_{1}}{dR}\\
+3y\left(  1+3y+y^{2}\right)  \left(  \frac{3}{R_{1}}\frac{\left(  \rho
_{0}-\rho_{\infty}\right)  }{\left(  1-y^{3}\right)  }\frac{dR_{1}}{d\rho_{0}%
}\right) \\
+2y^{2}\left(  1+3y+6y^{2}+5y^{3}\right)  \frac{dR_{1}}{dR}\left(  \frac
{3}{R_{1}}\frac{\left(  \rho_{0}-\rho_{\infty}\right)  }{\left(
1-y^{3}\right)  }\frac{dR_{1}}{d\rho_{0}}\right)
\end{array}
\right)  \allowbreak\nonumber\\
g_{\rho_{0}\rho_{0}}  &  =\frac{4\pi}{45}\frac{R^{5}}{\rho_{0}}+\frac{4\pi
}{45}\frac{1}{\left(  \rho_{\infty}-\rho_{0}y^{3}\right)  }R^{5}\frac{\left(
1-y\right)  ^{2}}{1+y+y^{2}}\left(
\begin{array}
[c]{c}%
\left(  5+6y+3y^{2}+y^{3}\right) \\
+3y\left(  1+3y+y^{2}\right)  \left(  \frac{3}{R_{1}}\frac{\left(  \rho
_{0}-\rho_{\infty}\right)  }{\left(  1-y^{3}\right)  }\frac{dR_{1}}{d\rho_{0}%
}\right) \\
+y\left(  1+3y+6y^{2}+5y^{3}\right)  \left(  \frac{3}{R_{1}}\frac{\left(
\rho_{0}-\rho_{\infty}\right)  }{\left(  1-y^{3}\right)  }\frac{dR_{1}}%
{d\rho_{0}}\right)  ^{2}%
\end{array}
\right)  \allowbreak\allowbreak\allowbreak\allowbreak\allowbreak
\allowbreak\allowbreak\nonumber
\end{align}
where $y \equiv R/R_{1}(R,\rho)$. Assuming that $R_{1}(R,0) > 0$ and that both
$\partial R_{1}/\partial R$ and $(\rho_{0}-\rho_{\infty})\partial R_{1} /
\partial\rho_{0}$ are well-behaved as $\rho_{0} \rightarrow\rho_{\infty}$, the
leading order density dependence of the kinetic coefficients becomes
\begin{align}
\label{gderiv}g_{RR}  &  \sim(\rho_{0} - \rho_{\infty})^{2}\\
g_{R\rho}  &  \sim(\rho_{0} - \rho_{\infty})\nonumber\\
g_{\rho_{0}\rho_{0}}  &  \sim1\nonumber
\end{align}
so that $\det g \sim(\rho_{0} - \rho_{\infty})^{2}$ just as in the naive model
discussed in the main text. This means that the only way to avoid the
divergence of the normalization of the probability density is if the range of
$R$ is finite, at least for small density fluctuations.

For the model used in the text, $R_{M}^{3} = \Delta R_{10}^{3}+\lambda\left(
\frac{\rho_{0}-\rho_{\infty}}{\rho_{\infty}}\right)  ^{2}R^{3}$, we have that
\begin{align}
\frac{\partial R_{1}}{\partial R} = 3\lambda\left(  \frac{\rho_{0}%
-\rho_{\infty}}{\rho_{\infty}}\right)  ^{2}R^{3}\\
\frac{\partial R_{1}}{\partial\rho_{0}} = 2\lambda\left(  \frac{\rho_{0}%
-\rho_{\infty}}{\rho^{2}_{\infty}}\right)  R^{3}\nonumber
\end{align}
Now, the leading order behavior of the coefficients is
\begin{align}
g_{RR}  &  =\frac{4\pi}{5}\frac{\left(  \rho_{0}-\rho_{\infty}\right)  ^{2}%
}{\rho_{\infty}}R^{3}\frac{1}{\left(  1-y\right)  \left(  1+y+y^{2}\right)
^{4}}\left(  5+6y+3y^{2}+y^{3}\right) \\
g_{R\rho_{0}}  &  =\frac{4\pi}{15}\frac{\left(  \rho_{0}-\rho_{\infty}\right)
}{\rho_{\infty}}R^{4}\frac{1}{\left(  1+y+y^{2}\right)  ^{3}}\left(
5+6y+3y^{2}+y^{3}\right) \nonumber\\
g_{\rho_{0}\rho_{0}}  &  =\frac{4\pi}{45}\frac{R^{5}}{\rho_{\infty}}%
+\frac{4\pi}{45}\frac{1}{\rho_{\infty} }R^{5}\frac{1-y}{\left(  1+y+y^{2}%
\right)  ^{2}}\left(  5+6y+3y^{2}+y^{3}\right) \nonumber
\end{align}
and
\begin{equation}
\det g = \left(  \frac{4\pi}{15}\frac{ \rho_{0}-\rho_{\infty}}{\rho_{\infty}%
}R^{4}\right)  ^{2} \frac{5+6y+3y^{2}+y^{3}}{\left(  1-y\right)  \left(
1+y+y^{2}\right)  ^{4}}%
\end{equation}

\section{The amplitude of the noise}

\label{MatrixQ}

The relation between the kinetic coefficients and the amplitude of the noise
is
\begin{equation}
g^{ij}=q^{ia}q^{jb}\delta_{ab}%
\end{equation}
Writing%
\begin{equation}
q^{ij}=\left(
\begin{array}
[c]{cc}%
a & b\\
c & d
\end{array}
\right)
\end{equation}
we need that%
\begin{align}
a^{2}+b^{2}  &  =g^{11}\\
ac+bd  &  =g^{12}\nonumber\\
c^{2}+d^{2}  &  =g^{22}\nonumber
\end{align}
Since there are more parameters than constraints, we are free to fix one
parameter. Choosing $c=0$ gives%
\begin{align}
a^{2}  &  =\frac{g^{11}g^{22}-\left(  g^{12}\right)  ^{2}}{g^{22}}\\
b^{2}  &  =\frac{\left(  g^{12}\right)  ^{2}}{g^{22}}\nonumber\\
d^{2}  &  =g^{22}\nonumber
\end{align}

\section{Numerical Integration}

\label{Simulations}

\subsection{Milstein scheme}

The Ito SDE is%
\begin{equation}
\frac{dx^{i}}{dt}=-Dg^{ij}\frac{\partial\beta\Omega}{\partial x^{j}%
}+D\varepsilon^{2}A^{i}+\varepsilon\sqrt{2D}q_{a}^{i}\xi^{a}\left(  t\right)
\end{equation}
Setting $D=1$ by an appropriate choice of time scale, strong first-order
convergence can be achieved with the Milstein scheme\cite{Kloeden}%
\begin{equation}
x^{i}\left(  t+\tau\right)  =x^{i}\left(  t\right)  -\left[  g^{ij}%
\frac{\partial\beta\Omega}{\partial x^{j}}+\varepsilon^{2}A^{i}\right]
_{\mathbf{x}\left(  t\right)  }\tau+2q_{a}^{j}\frac{\partial q_{b}^{i}%
}{\partial x^{j}}I^{ab}\tau+\varepsilon\sqrt{2}q_{a}^{i}\left(  \mathbf{x}%
\left(  t\right)  \right)  I^{a}\sqrt{\tau}%
\end{equation}
where%
\begin{equation}
I^{a}=\frac{1}{\sqrt{\tau}}\int_{t}^{t+\tau}dW_{t^{\prime}}^{a}=\frac{1}%
{\sqrt{\tau}}\left(  W_{t+\tau}-W_{t}\right)  =\xi_{t}\in\mathcal{N}\left(
0,1\right)
\end{equation}
and the Levy area is%
\begin{equation}
I^{ab}=\frac{1}{\tau}\int_{t}^{t+\tau}dW_{t^{\prime}}^{a}dW_{t^{\prime}}^{b}%
\end{equation}
In particular,
\begin{equation}
I^{aa}=\frac{1}{2}\left(  \left(  I^{a}\right)  ^{2}-1\right)
\end{equation}
while the off-diagonal terms can be calculated from\cite{Kloeden}%
\begin{equation}
I^{ab}\simeq\frac{1}{2}I^{a}I^{b}+\sqrt{\rho_{p}}\left(  \mu^{a}I^{b}-\mu
^{b}I^{a}\right)  +\frac{1}{2\pi}\sum_{r=1}^{p}\frac{1}{r}\left\{  \zeta
_{r}^{a}\left(  \sqrt{2}I^{b}+\eta^{b}\right)  -\zeta_{r}^{b}\left(  \sqrt
{2}I^{a}+\eta^{a}\right)  \right\}
\end{equation}
with%
\begin{equation}
\mu^{a},\zeta_{r}^{a},\eta^{a}\in\mathcal{N}\left(  0,1\right)
\end{equation}
and the constant%
\begin{equation}
\rho_{p}=\frac{1}{12}-\frac{1}{2\pi^{2}}\sum_{r=1}^{p}\frac{1}{r^{2}}%
\end{equation}
for some fixed value of $p$ (typically, $p=5$ or $p=10$\cite{SDE1,SDE2}).

\subsection{Variable time step}

When we detect that the integration over a timestep $\delta t$ is too
inaccurate (see below), the timestep is halved and an extra point is added.
The correct way to do this is via the ``Brownian Bridge'' technique\cite{SDE0}
(see also \cite{SDE1,SDE2}), whereby the difference
\begin{equation}
W_{n+1}^{\alpha}-W_{n}^{\alpha}=\sqrt{\Delta t_{n}}\xi_{n},\;\xi_{2n}%
\in\mathcal{N}\left(  0,1\right)
\end{equation}
is replaced by two stochastic steps%
\begin{align}
W_{2n+1}^{\alpha}-W_{2n}^{\alpha}  &  =\frac{1}{2}\left(  W_{n+1}^{\alpha
}-W_{n}^{\alpha}\right)  +\sqrt{\frac{\Delta t_{n+1}}{2}}\xi_{2n+1}%
=\sqrt{\Delta t_{n}}\left(  \frac{1}{2}\xi_{n}+\frac{1}{2}\xi_{2n+1}\right) \\
W_{2n+2}^{\alpha}-W_{2n+1}^{\alpha}  &  =\frac{1}{2}\left(  W_{n+1}^{\alpha
}-W_{n}^{\alpha}\right)  -\sqrt{\frac{\Delta t_{n+1}}{2}}\xi_{2n+1}%
=\sqrt{\Delta t_{n}}\left(  \frac{1}{2}\xi_{n}-\frac{1}{2}\xi_{2n+1}\right)
\nonumber
\end{align}

Thus, we replace the Gaussian deviates%
\begin{equation}
I_{n}^{\alpha}\rightarrow\left\{
\begin{array}
[c]{c}%
\widetilde{I}_{n}^{\alpha}=\frac{1}{2}I_{n}^{\alpha}+\frac{1}{2}I_{n1}%
^{\alpha}\\
\widetilde{R}_{n+1/2}^{\alpha}=\frac{1}{2}I_{n}^{\alpha}-\frac{1}{2}%
I_{n1}^{\alpha}%
\end{array}
\right.  ,\;\;R_{n1}^{\alpha}\in\mathcal{N}\left(  0,1\right)
\end{equation}
and%
\begin{align}
x_{n+1/2}^{i}  &  =x_{n}^{i}+\left(  \frac{\delta t}{2}\right)  A^{i}\left(
\mathbf{x}_{n}\right)  +\sqrt{2\left(  \delta t\right)  }q_{\alpha}^{i}\left(
\mathbf{x}_{n}\right)  \widetilde{I}_{n}^{\alpha}+2\left(  \frac{\delta t}%
{2}\right)  q_{\alpha}^{j}\left(  \mathbf{x}_{n}\right)  \frac{\partial
q_{\beta}^{i}\left(  \mathbf{x}_{n}\right)  }{\partial x_{n}^{j}}%
I^{\alpha\beta}\left(  t,t+\frac{\delta t}{2};\widetilde{\mathbf{R}}%
_{n}\right) \\
x_{n+1}^{i}  &  =x_{n+1/2}^{i}+\left(  \frac{\delta t}{2}\right)  A^{i}\left(
\mathbf{x}_{n+1/2}\right)  +\sqrt{2\left(  \delta t\right)  }q_{\alpha}%
^{i}\left(  \mathbf{x}_{n+1/2}\right)  \widetilde{I}_{n+1/2}^{\alpha
}\nonumber\\
&  +2\left(  \frac{\delta t}{2}\right)  q_{\alpha}^{j}\left(  \mathbf{x}%
_{n+1/2}\right)  \frac{\partial q\left(  \mathbf{x}_{n+1/2}\right)  }{\partial
x_{n+1/2}^{j}}I^{\alpha\beta}\left(  t+\frac{\delta t}{2},t+\delta
t;\widetilde{\mathbf{R}}_{n+1/2}\right) \nonumber
\end{align}
In practice, it is more convenient to replace all of the $\delta t$'s by half
their value so we use the equivalent formulation%
\begin{equation}
R_{n}^{\alpha}\rightarrow\left\{
\begin{array}
[c]{c}%
\widetilde{R}_{n}^{\alpha}=\frac{1}{\sqrt{2}}R_{n}^{\alpha}+\frac{1}{\sqrt{2}%
}R_{n1}^{\alpha}\\
\widetilde{R}_{n+1/2}^{\alpha}=\frac{1}{\sqrt{2}}R_{n}^{\alpha}-\frac{1}%
{\sqrt{2}}R_{n1}^{\alpha}%
\end{array}
\right.  ,\;\;R_{n1}^{\alpha}\in\mathcal{N}\left(  0,1\right)
\end{equation}
and%
\begin{align}
x_{n+1/2}^{i}  &  =x_{n}^{i}+\left(  \frac{\delta t}{2}\right)  A^{i}\left(
\mathbf{x}_{n}\right)  +\sqrt{2\left(  \frac{\delta t}{2}\right)  }q_{\alpha
}^{i}\left(  \mathbf{x}_{n}\right)  \widetilde{R}_{n}^{\alpha}\\
&  +2\left(  \frac{\delta t}{2}\right)  q_{\alpha}^{j}\left(  \mathbf{x}%
_{n}\right)  \frac{\partial q_{\beta}^{i}\left(  \mathbf{x}_{n}\right)
}{\partial x_{n}^{j}}I^{\alpha\beta}\left(  t,t+\frac{\delta t}{2}%
;\widetilde{\mathbf{R}}_{n}\right) \nonumber\\
x_{n+1}^{i}  &  =x_{n+1/2}^{i}+\left(  \frac{\delta t}{2}\right)  A^{i}\left(
\mathbf{x}_{n+1/2}\right)  +\sqrt{2\left(  \frac{\delta t}{2}\right)
}q_{\alpha}^{i}\left(  \mathbf{x}_{n+1/2}\right)  \widetilde{R}_{n+1/2}%
^{\alpha}\nonumber\\
&  +2\left(  \frac{\delta t}{2}\right)  q_{\alpha}^{j}\left(  \mathbf{x}%
_{n+1/2}\right)  \frac{\partial q\left(  \mathbf{x}_{n+1/2}\right)  }{\partial
x_{n+1/2}^{j}}I^{\alpha\beta}\left(  t+\frac{\delta t}{2},t+\delta
t;\widetilde{\mathbf{R}}_{n+1/2}\right) \nonumber
\end{align}

\subsection{Quality Control}

A variable time step scheme is desirable due to the fact that both the
thermodynamic forces and the kinetic coefficients can be anywhere between zero
and divergent in magnitude. In order to provide a quick and easy assessment of
the accuracy of the integration scheme, we have two criterion for reducing the
time step. The first is if the result of advancing one of the parameters
causes it to assume an unphysical value (e.g. the radius or density are less
than zero). The thermodynamic forces and kinetic coefficients will in general
diverge as these limits are reached thus preventing escape from the physical
region but if the time step is too large, the integrator may miss such
divergences so reducing the time step is obviously needed.

The second criterion, when all values are physical, is that the distance moved
in parameter space is less than some prescribed value, $\Delta s$. The
stochastic process itself imposes a Riemannian geometry with a precisely
defined measure of distance in parameter space, see e.g. Ref.
\onlinecite{Lutsko_JCP_2012_1}. In principle, given a path defined by $R(t),
\rho_{0}(t)$ for some beginning and ending times, $t_{0},t_{1}$, the distance
moved, $s$, is
\begin{equation}
s = \int_{t_{0}}^{t_{1}} \left(  \sqrt{ \frac{dx^{i}(t)}{dt}g_{ij}%
(x(t))\frac{dx^{j}(t)}{dt}} \right)  dt
\end{equation}
Rather than evaluate this expression exactly, which would be somewhat
expensive, we use the approximation
\begin{equation}
s^{(approx)} = \left(  x^{i}(t_{1}) - x^{i}(t_{0}) \right)  (\frac
{g_{ij}(t_{0}) + g_{ij}(t_{1})}{2}\left(  x^{j}(t_{1}) - x^{j}(t_{0}) \right)
\end{equation}
and we reduce the time step whenever $s^{(approx)} > \Delta s$. The results in
the main text were obtained for $\Delta s^{*} = 0.05$. \bigskip

\section{Calculation of average transition paths}

\label{AveragePaths} The average pathway from the FFS simulations is
determined by the average crossing point of each intermediate surface. For
each surface, the simulations provide a population of $N$ crossing points each
of which is specified by a set of coordinates (density and radius) which we
denote as $(R_{\alpha},\rho_{\alpha})$. Clearly, one cannot simply average the
coordinates as this would generally not even produce a point on the surface.
Instead, let the curve be parameterized as $(R(u),\rho(u)$ with $0 \le u \le
u_{max}$. Then, the i-th crossing point will correspond to some value
$u_{\alpha}$ such that $(R_{\alpha},\rho_{\alpha}) = (R(u_{\alpha}%
),\rho(u_{\alpha}))$. We cannot simply average the values of $u$ since the
result would then depend on the chosen parameterization of the curve (for
which there are an unlimited number of in-equivalent possibilities). So, for
each point we calculate the distance along the surface (which is simply a line
in our two-dimensional case) using the natural metric provided by the matrix
$g_{ij}$. Specifically,
\begin{equation}
d_{\alpha} = \int_{0}^{u_{\alpha}} \sqrt{\frac{d x^{i}(u)}{du}g_{ij}%
(\mathbf{x}(u))\frac{d x^{j}(u)}{du}} du
\end{equation}
where, as previously, $x^{0} \equiv R$ and $x^{1} \equiv\rho$ and repeated
indices are summed. The distances along the curve as calculated here are
independent of any reparameterization of the curve and, as well, are covariant
with respect to a change of the variables characterizing the cluster. It is
therefore these quantities that we average to get the average value of $d$ and
from this the coordinates of the average point of crossing.
\end{document}